\documentclass[fleqn,usenatbib]{mnras}
\usepackage{newtxtext,newtxmath}
\usepackage[T1]{fontenc}
\usepackage{ae,aecompl}
\usepackage{graphicx}	% Including figure files
\usepackage{amsmath}	% Advanced maths commands
\usepackage{times}

\usepackage[normalem]{ulem}

\defcitealias{Asad20}{Paper~I}
\defcitealias{Goudfrooij21}{Paper~II}
\defcitealias{Asad21}{Paper~III}

\title[Star clusters: full-spectrum fitting vs.\ colour-magnitude diagram fitting]
{On the precision of full-spectrum fitting of simple stellar populations. IV. A systematic comparison with results from colour-magnitude diagrams.}

\author[Asa'd et al.]{Randa Asa'd,$^{1,2}$\thanks{E-mail: raasad@aus.edu}
Paul Goudfrooij,$^{2}$
and A. M. As'ad$^{3}$\\
$^{1}$American University of Sharjah, Physics Department, P.O.Box 26666, Sharjah, UAE\\
$^2$Space Telescope Science Institute, 3700 San Martin Drive, Baltimore, MD 21218, USA\\
$^3$King Abdullah II School for Information Technology, University of Jordan, Amman, Jordan
}

\date{Accepted 2022 February 23. Received 2022 February 23; in 
  original form 2021 November 24}
\pubyear{2022}

\begin{document}

\label{firstpage}
\pagerange{\pageref{firstpage}--\pageref{lastpage}}
\maketitle

\begin{abstract}
In this fourth paper of a series on the precision of ages of stellar populations obtained through the full-spectrum fitting technique, we present a first systematic analysis that compare the age, metallicity and reddening of star clusters obtained from resolved and unresolved data (namely colour-magnitude diagrams (CMDs) and integrated-light spectroscopy) using the same sets of isochrones. We investigate the results obtained with both Padova isochrones and MIST isochrones.
We find that there generally is a good agreement between the ages derived from CMDs and integrated spectra. However, for metallicity and reddening, the agreement between results from analyses of CMD and integrated spectra is significantly worse.
Our results also show that the ages derived with Padova isochrones match those derived using MIST isochrones, both with the full spectrum fitting technique and the CMD fitting method. However, the metallicity derived using Padova isochrones does not match that derived using MIST isochrones using the CMD method.
We examine the ability of the full-spectrum fitting technique in detecting age spreads in clusters that feature the extended Main Sequence Turn Off (eMSTO) phenomenon using two-population fits. We find that 3 out of 5 eMSTO clusters in our sample are best fit with one single age, suggesting that eMSTOs do not necessarily translate to detectable age spreads in integrated-light studies.

\end{abstract}

 \begin{keywords}
galaxies: star clusters: general -- stars: massive
\end{keywords}

 \section{Introduction}

The determination of accurate ages and metallicities of star clusters enables one to study the formation and evolution of stellar systems. 
While much of our knowledge on ages and metallicities of star clusters in the Local Group is based on photometry and spectroscopy of spatially resolved data, observations of more distant stellar systems have to rely on integrated-light observations.  It is therefore important to determine the accuracy and precision of population properties derived from integrated-light spectroscopy by using star clusters for which precise ages and metallicities have already been determined from resolved data such as colour-magnitude diagrams (CMDs).
Some efforts towards this goal have already been made in the past. For example,  \cite{Garcia12} studied the metallicity and SFH of a blue compact dwarf galaxy from both CMD and integrated spectroscopy. \cite{Ruiz-Lara15, Ruiz-Lara18} studied the SFH recovery using both resolved and integrated-light data for the Large Magellanic Cloud (LMC) and Leo A, respectively, and several old globular clusters in the Milky Way have been investigated in this manner \citep[e.g.,][]{Gibson99, Puzia02, Kuncarayakti16, Goncalves20}. 

In the previous papers of this series we provided a detailed theoretical ground for the precision of the full spectrum fitting technique in determining the age and metallicity of stellar populations. In  \citet[][hereafter Paper I]{Asad20}, we investigated the precision of the ages and metallicities of 21,000 mock simple stellar populations (SSPs) determined through full-spectrum fitting in the optical range. In  \citet[][Paper II]{Goudfrooij21} we studied the influence of star cluster mass through stochastic fluctuations of the number of stars near the top of the stellar mass function as functions of age and wavelength regime (from the blue optical through the mid-infrared).  In \citet[][Paper III]{Asad21} we examined the precision of the full-spectrum fitting method in deriving possible age spreads within a star cluster using 118,800 mock star clusters. In this work we investigate the precision of the full-spectrum fitting technique in deriving the age and metallicity of five star clusters in the LMC for which both integrated-light spectra and deep CMDs are available, and we discuss the results in light of the findings of our previous theoretical studies.

Precision studies of real observations are tricky because they require other independent studies that derive the same parameters to be used for the comparison. In order to assess the precision of ages and metallicities obtained using integrated-light spectra one can compare the results with those obtained from spatially resolved photometry when available. However, it should be noted that ages obtained from resolved data by fitting the isochrones on the color-magnitude diagrams (CMDs) of star clusters often differ from one study to another due to the use of different isochrone sets as well as  different approaches in using the isochrones (e.g., using isochrones of fixed metallicities versus fitting for both metallicity and age).  This can lead to deriving different ages for a star cluster using \textit{the same} observed data.
Using a sample of 7 intermediate-age clusters in the LMC,  \citet{Goudfrooij11a} quantified systematic uncertainties in derived age, [Fe/H], distance, and reddening from CMDs by comparing the best-fit results from the Padova, Teramo (BaSTI) and Dartmouth isochrone sets. Their results yield systematic uncertainties of $\pm$7\% in age, $\pm$0.1 dex in [Fe/H], $\pm$0.05 mag in $(m\!-\!M)_{0}$ ($\simeq$ 5\% in linear distance), and $\pm$0.02 mag in $A_V$  ($\simeq$15\%).
In this work we aim to avoid uncertainties associated with the details of isochrone fitting for deriving the ages of our sample. As such, we focus our study on an observed sample of five star clusters in the LMC that has \textit{both} observed integrated spectra as well as resolved photometric data to be able to compare the results in a systematic way.

All star clusters in our sample exhibit the extended main sequence turnoff (eMSTO) phenomenon, where the MSTO is significantly wider than can be explained by photometric uncertainties or stellar binarity.
The eMSTO has been observed in young (>20 Myr, e.g., \citep{Milone15, Bastian16}) through intermediate-age ($\la$\,2 Gyr, e.g. \citep{Milone09, Correnti14, Goudfrooij2014, Goudfrooij2015, Niederhofer2016a}) massive star clusters of the Magellanic Clouds (MCs).
Two major scenarios have been proposed for the origin of this feature. Originally, the eMSTO was proposed to be caused by multiple star forming events during a period of a few $10^8$ yr within the cluster (i.e., age spreads, \citep{Mackey07,Milone09,Goudfrooij11b,Goudfrooij2014,Correnti14}). However, this scenario is not without drawbacks. One of these is that massive young clusters with ages $\sim 10$ Myr or older have not been observed to host secondary star formation events \citep{Bastian13b,Cabrera-Ziri16}. Additionally, this scenario cannot explain the strong correlation found by \citet{Niederhofer15a} between the inferred age spread and the age of the cluster.

\citet{Bastian09} proposed an alternative scenario for the origin of the eMSTO, namely that this was caused by a range of stellar rotation rates in a single-aged population. They argued that rotation lowers the luminosity and effective temperature at the stellar surface, which could cause an eMSTO depending on the distribution of viewing angles and rotation rates. However, subsequent studies found an opposite (and stronger) effect of rotation, namely a longer stellar lifetime due to internal mixing, thus yielding a brighter and bluer star \citep{Girardi11,Ekstrom12}. After comprehensive isochrone models that include the effects of rotation became available \citep{Georgy14,Girardi19,Costa19}, it became clear that many properties of eMSTOs can be explained by stellar rotation, including the aforementioned age spread vs.\ cluster age relation \citep[but see][]{Goudfrooij17,Goudfrooij18b}. Also, in this scenario, clusters older than 2 Gyr are not expected to host eMSTOs, as stars on the turnoff in 2 Gyr old clusters are not expected to be strong rotators, due to the presence of magnetic fields capable of braking the star's rotation \citep{Brandt15}. Indeed, \cite{Martocchia18} found that the 2 Gyr-old cluster NGC 1978 does not show a prominent eMSTO, consistent with predictions from the stellar rotation scenario.

Irrespective of the nature of the eMSTO, there is a current lack of coverage of this phenomenon using integrated-light spectra in the literature. In particular, it is unclear whether the eMSTO feature has a significant affect on the age (and/or the inferred age spread) derived from integrated-light spectra.  While there have been many studies looking into integrated spectroscopy of young, intermediate age, and old star clusters in distant galaxies \citep[e.g.,][]{Puzia05,Trancho07,Asad14, Asad13, Asad16, Asad17, Cabrera-Ziri14}, the Magellanic Clouds offer a unique opportunity, due to the presence of massive clusters with a range of ages that can be studied through integrated spectroscopy as well as resolved stellar photometry. \cite{Cabrera-Ziri14, Cabrera-Ziri16} studied the integrated light spectra of two very massive clusters (NGC 34-1 and NGC 7252-W3) with ages between 100-500~Myr in order to search for evidence of multiple star-formation events within the clusters.  The authors find that both clusters are well described by an SSP, with no evidence for multiple epochs of star-formation.  However, as both clusters are unresolved, it is not known whether they display the eMSTO phenomenon.

In \citetalias{Asad21}, the precision of the full spectrum fitting in deriving possible age spreads within a star cluster was examined on 118,800 mock star clusters covering all ages in the range 6.8 $<$ log\,(age/yr) $<$ 10.2, with mass fractions from 10\% to 90\% for two age gaps (0.2 dex and 0.5 dex). Random noise was added to the model spectra to achieve S/N ratios between 50 to 100 per wavelength pixel.
It was found that the mean of the derived Age\,(SSP1) generally matches the real Age\,(SSP1) to within 0.1 dex up to ages around log (age/yr) = 9.5. The precision decreases for log (age/yr) $>$ 9.6 for any mass fraction or S/N, due to the similarity of SED shapes for those ages. For the same reason, the uncertainty of the mean of the derived Age\,(SSP2) is higher than that of Age\,(SSP1). Increasing the age gap in the mock clusters improves the derived parameters, but Age\,(SSP2) is still overestimated for the young ages. \\
In this work we will use our sample of star clusters to compare the predicted age gap from integrated spectra with that obtained from CMDs.

In order to investigate the precision of the full-spectrum fitting technique in deriving the age and metallicity of a sample of star clusters with both integrated-light spectra and deep colour-magnitude diagrams (CMDs), we first discuss the sample selection and description in Section \ref{Literature}, then do a complete systematic CMD analysis of our sample in Section \ref{s:CMDanalysis}. After that we perform the full-spectrum fitting of our sample in Section \ref{s:spectral_analysis}. We summarize our finidngs in Section \ref{Summary}.

%In section 2 we summarize the literature findings for five young and intermediate-age resolved LMC star clusters (NGC\,1651, NGC\,1850, NGC\,1863, NGC\,2173 and NGC\,2213). In section 3 we use the resolved data of our sample to derive new consistent ages using two isochrone sets. In section 4 we derive the ages using the full spectrum fitting technique and compare the results with the ones obtained from CMDs. In section 5 we shed light on results of possible age spreads obtained using the two techniques. The conclusion is presented in section 5.

\section{Sample Selection and Description}
\label{Literature}

Our goal is to perform a systematic analysis to compare the age, metallicity and reddening of star clusters derived from CMDs with those derived from integrated-light spectroscopy. To achieve this, we select a sample of star clusters for which both high-quality integrated-light spectra and resolved HST photometry are available. Based on the data availability our final sample has five star clusters (NGC\,1651, NGC\,1850, NGC\,2173, NGC\,2213, and NGC\,2249; see Table \ref{T-Literature} for their relevant properties).

We start by discussing the main findings of the photometric studies of the clusters in our sample from the literature. 

 \subsection{NGC 1651}

\citet{Li2014} reported an eMSTO and a very narrow subgiant branch (SGB) for NGC 1651, by using archival Hubble Space Telescope (HST) data. They suggested that the narrowness of the SGB was inconsistent with a range of ages, and that stellar rotation could be an explanation. However,  \citet{Goudfrooij2015} found that when the effects of unresolved binaries and stochastic effects are taken into account, the SGB morphology of NGC~1651 is actually consistent with the age distribution derived by \citet{Goudfrooij2014}.
%using new HST data noticed an eMSTO, and proposed that it is due to age spread, not stellar rotation.
%PG I changed this paragraph, because that's not exactly what I concluded in that 2015 paper. I clarified why the analysis of Li+2014 was incorrect and that the morphology of the SGB was actually  consistent with a range of ages.

\citet{Li2015} further analyzed NGC 1651 and they tested different models from which they conclude that the best fit comes when using 50\% binaries and 70\% rotation in addition to an age spread from 1.8 Gyr to 1.4 Gyr ago. Their investigation suggests that a 1.5 Gyr SSP with rotation can also fit well. Using the same data of \citet{Goudfrooij2014}, \citet{Niederhofer2016a} compared the distributions of stars that populate the SGB and the red clump. They could not confirm nor deny large age spreads in intermediate-age LMC clusters (including NGC\,1651) due to uncertainties.

 \subsection{NGC 1850}

\citet{Niederhofer15a} analyzed the color-magnitude diagrams of NGC\,1850 and fitted its star formation history to derive upper limits of potential age spreads. They showed that despite having properties similar to intermediate-age clusters ($1-2$~Gyr), NGC 1850 did not show an age spread of a few hundred Myr or more (i.e. similar to what had been inferred for the $1-2$~Gyr clusters).  Instead, an upper limit to the age spread of $\sim40$~Myr could be placed.

\citet{Bastian16} analyzed HST photometry of NGC 1850 and found evidence for splitting in the main sequence and eMSTO. They found that the use of non-rotating stellar isochrones leads to an age spread of  about 40 Myr. However, they did not find evidence for multiple, isolated episodes of star-formation bursts within the cluster. They suggested that the cause of the eMSTO could be due to stellar rotation. Additionally, \citet{Bastian17b} detected a large number of Be stars in the MSTO of NGC 1850, which supports the presence of a large population of rapidly rotating stars within the cluster, with the majority of stars rotating near the critical velocity.

\citet{Correnti17a} used HST WFC3 photometry in different filters than the ones used by \citet{Bastian16}. They demonstrated that the global CMD morphology is well-reproduced by a combination of SSPs that cover an age range of about 35 Myr as well as a wide variety of stellar rotation rates. They also confirm that the MSTO region hosts a population of Be stars, providing further evidence that rapidly rotating stars are present in the cluster.

\subsection{NGC 2173 and NGC 2213}

\citet{Bertelli03} found a wide MSTO in NGC 2173 and analyzed it using Padova models, suggesting an extended star formation period (spanning about 300 Myr), beginning 1.7 Gyr and ending 1.4 Gyr ago in NGC 2173.  \citet{Goudfrooij2014} found similar results for both NGC 2173 and NGC 2213, using deep data from {\it HST/WFC3}. 
%The derived age and age spread, however, turned out to be consistent with expectations from the {\sc syclist} isochrone models of \citet{Georgy14} that involve stellar rotation \citep[see][]{Niederhofer15b}.

Both clusters were part of the sample used in \citet{Piatti16} who used ground-based photometry results to argue that the clusters' core radii, masses, and dynamical state are not the cause of the eMSTOs. Based on this, they concluded that the eMSTO is not due to age spreads within the clusters.

These clusters were also part of the sample analyzed in \citet{Niederhofer2016a}.

\subsection{NGC 2249}

\citet{Piatti14} reported NGC 2249 as a new eMSTO cluster candidate on the basis of its dual red clump.
\citet{Correnti14} compared the observed CMD for this cluster with Monte Carlo simulations and found that the MSTO region is significantly wider than that derived from simulations of simple stellar populations. They suggested that the eMSTO morphology was due to a range of stellar ages. %rather than a range of stellar rotation velocities or interacting binaries.

\citet{Goudfrooij17} studied the details of differences in MSTO morphology expected from spreads in age versus spreads in rotation rates, using Monte Carlo simulations and compared the simulations with high-quality HST data for NGC 2249.
The authors found that NGC 2249 fits the age spread vs.\ age relation expected from stellar rotation.  However, they also found that the detailed morphology of the eMSTO was not well fit by any particular range of rotation speeds and viewing angles, and suggested a mixture of age spread and rotation to explain the observations.

This cluster was also part of the samples used in the
studies of \citet{Niederhofer15b} and \citet{Piatti16} mentioned above.

\begin{table}
\caption{Details of our sample}
\label{T-Literature}
\begin{tabular}{@{}lccc@{}}
\hline \hline
Cluster & log (age/yr) & Age Spread$^1$ & $\log ({\cal{M}}/M_{\odot})^2$ \\
\hline
NGC\,1651 & 9.3$^a$ & 315$^a$ & 4.67$^a$\\
NGC\,1850 & 8.0$^b$ & 44$^c$ &  4.62$^c$\\
%NGC\,2031 & 8.3 & 90 & $This Work$\\
NGC\,2173 & 9.2$^a$ & 431$^a$ & 4.43$^a$ \\
NGC\,2213 & 9.2$^a$ & 329$^a$ & 4.22$^a$ \\
NGC\,2249 & 9.0$^d$ & 235$^e$ & 4.24$^d$ \\
\hline
\multicolumn{4}{@{}l}{$^1$ in Myr.}\\
\multicolumn{4}{@{}l}{$^2$ Cluster masses assume a \citet{Kroupa01} IMF.}\\
\multicolumn{4}{@{}l}{$a$ \cite{Goudfrooij2014}}\\
\multicolumn{4}{@{}l}{$b , c$ \cite{Bastian16}, \citet{Correnti17a} }\\
\multicolumn{4}{@{}l}{$d , e$ \cite{Correnti14}, \cite{Goudfrooij17} }\\

\end{tabular}
\end{table}

\begin{figure*}
\includegraphics[width=11.7cm]{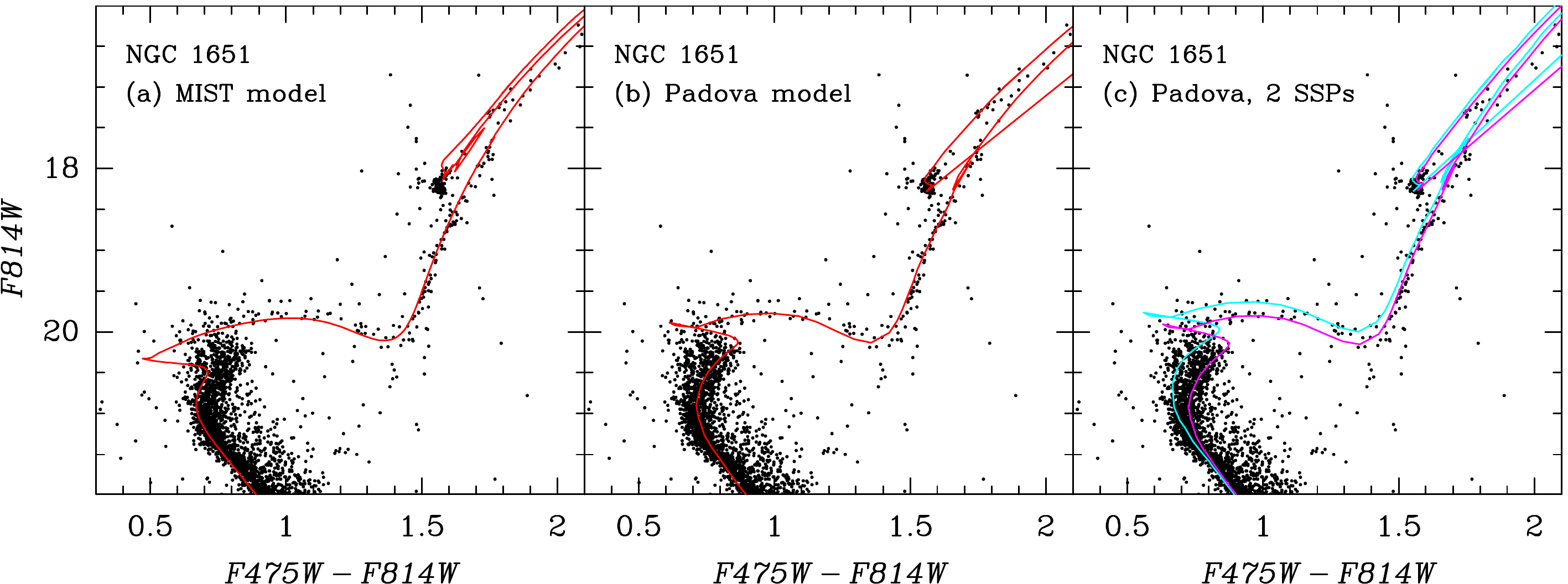} \\
\vspace*{1mm}
\includegraphics[width=11.7cm]{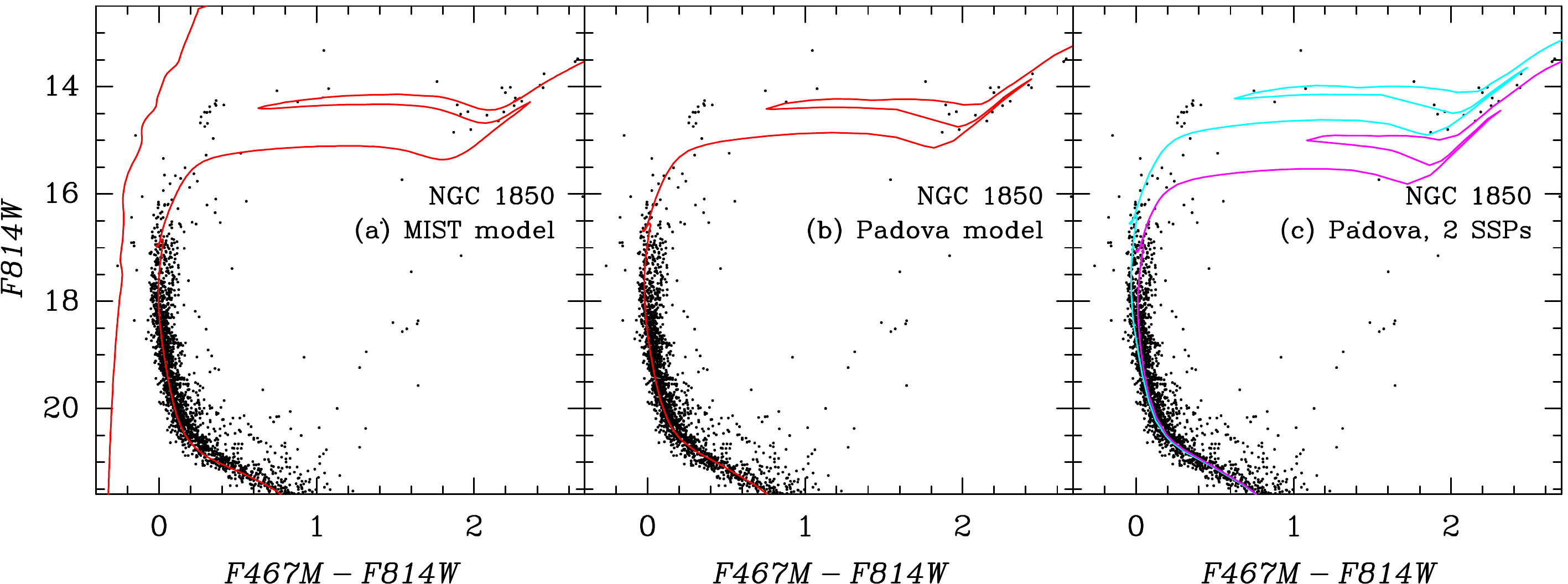} \\
\vspace*{1mm}
\includegraphics[width=11.7cm]{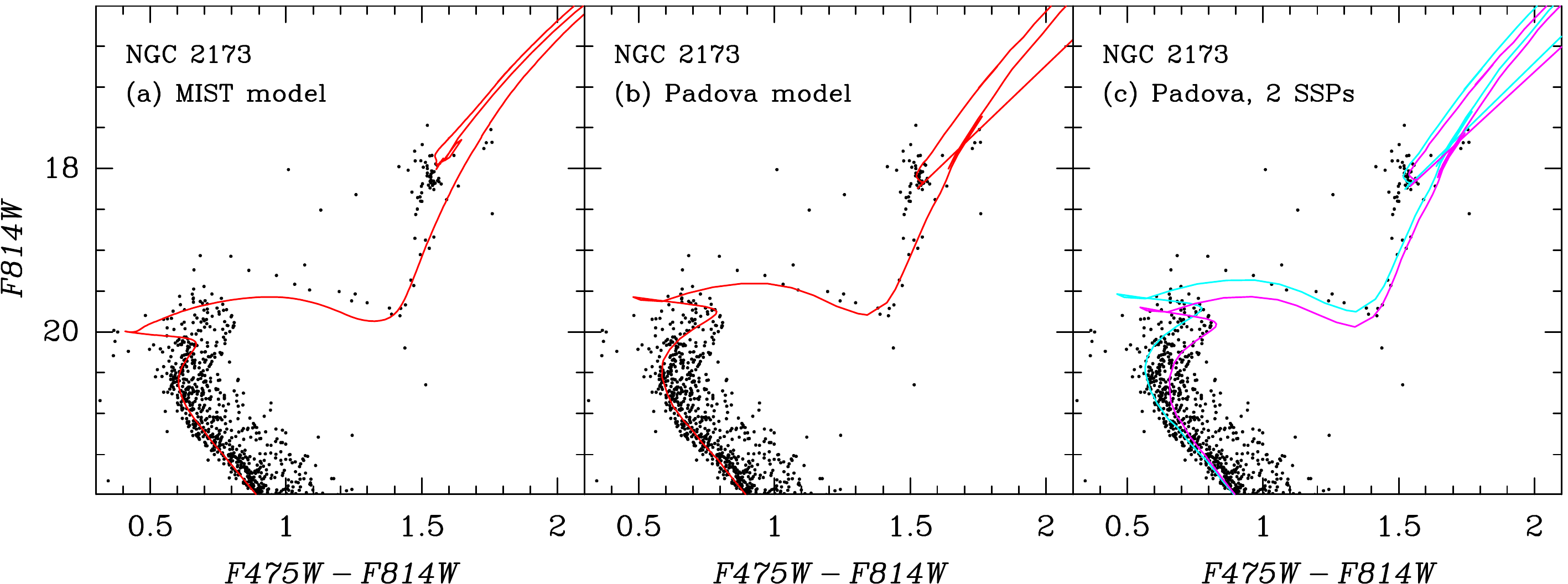} \\
\vspace*{1mm}
\includegraphics[width=11.7cm]{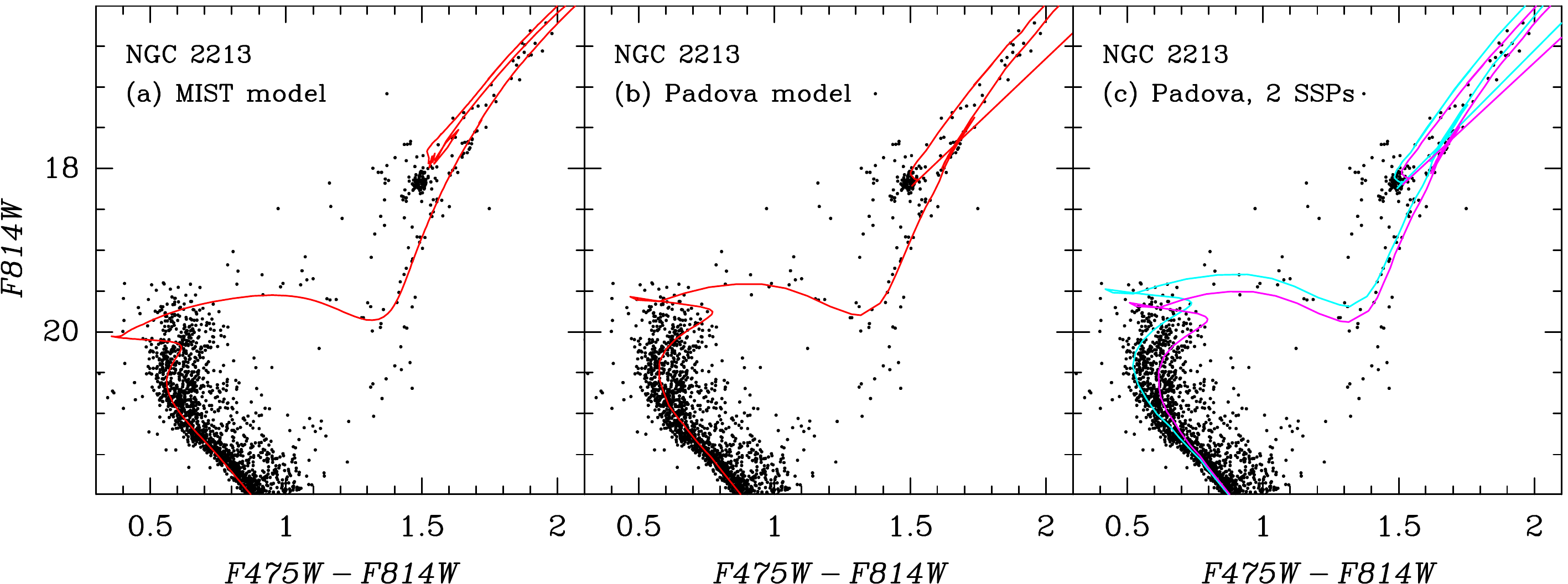} \\
\vspace*{1mm}
\includegraphics[width=11.7cm]{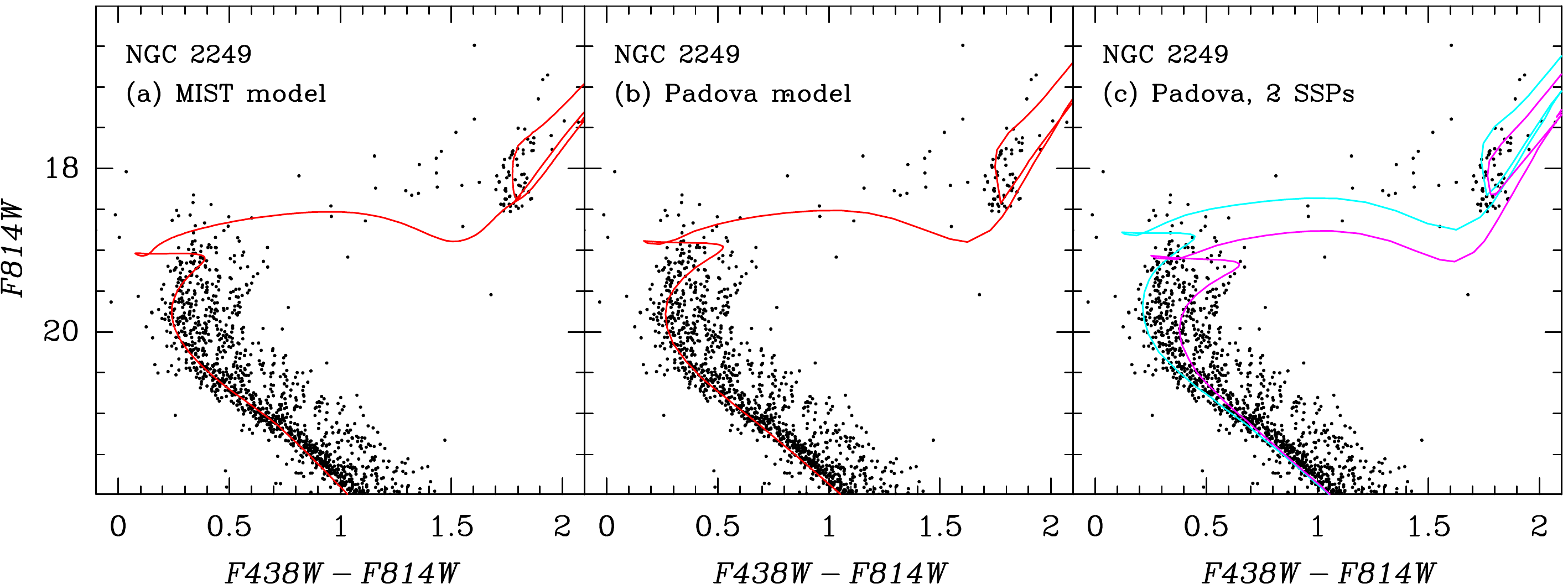}
\caption{CMDs from HST/WFC3 data of the clusters in our sample. The photometry data was taken from \citet{Goudfrooij2014} and  \citet{Correnti14,Correnti15}. The left panels show the best-fit MIST isochrone \citep{Choi16} in red, the central panels show the best-fit Padova isochrone \citep{Marigo08} in red, and the right-hand panels show the best-fit 2-SSP combination of Padova isochrones in light blue and magenta. Age, metallicity, reddening, and distance of all best-fit isochrones are listed in Table~\ref{All_CMD}. See text for details. }
\label{f:CMD_analysis}
\end{figure*}

\begin{figure*}
\resizebox{150mm}{!}{\includegraphics[angle=0]{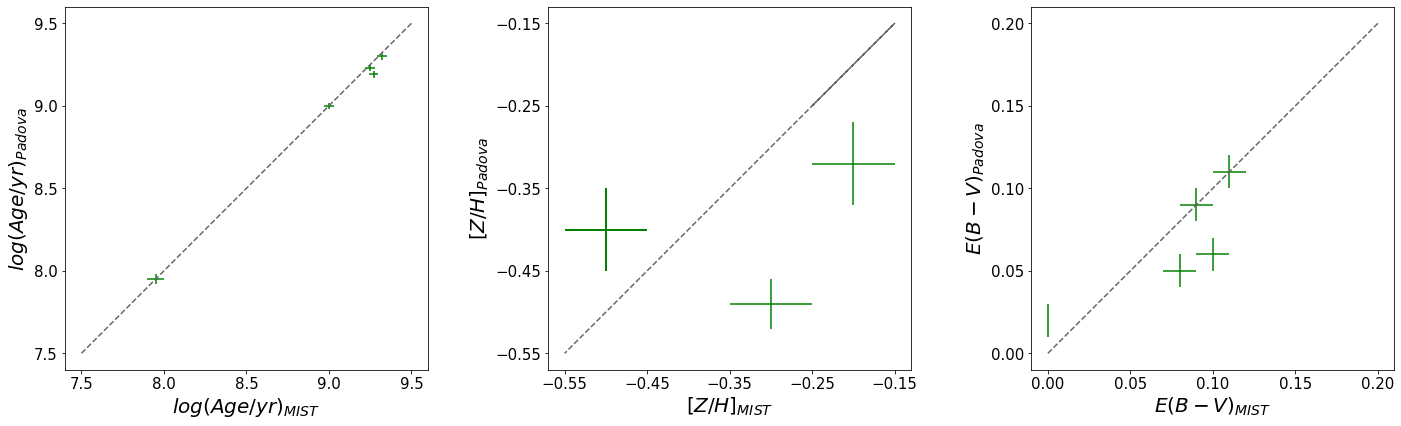}} \\
\caption{The correlation between the results derived using the CMD method with Padova and MIST models.}
\label{compare_CMD}
\end{figure*}

\section{Complete Systematic CMD ages of our sample}
\label{s:CMDanalysis}

To enable a systematic  apples-to-apples comparison between stellar population parameters derived from analysis of CMDs vs.\ of integrated-light spectra (see Section~\ref{s:spectral_analysis} below), we re-analyse the available \emph{HST} photometry of the clusters in our sample using the MIST and Padova isochrone models. Details on the \emph{HST} data used here are listed in Table~\ref{tab:HSTobs}.  Note that we only use non-rotating isochrones for this paper, since isochrone models that include effects of rotation only cover a rather limited range of stellar masses, which renders them incompatible with our intended goal to compare analyses of CMD data and integrated-light spectra.
Best-fit isochrones are selected using the method described in detail in \citet{Goudfrooij09,Goudfrooij11a}, which selects isochrones for which the population parameters provide a best fit to the relative brightnesses and colours of core helium-burning stars and the MSTO, and the brightness and color of the RGB bump (the latter only for clusters older than about 1.2 Gyr). After fitting the relative brightnesses and colours mentioned above, distance and reddening are then derived by means of least-squares fitting of the magnitudes and colours of the MSTO and red clump in the isochrone to those in the data.

\begin{table}
  \caption{Observational details on \emph{HST}/WFC3 data of sample clusters}
  \begin{tabular}{@{}lcllcl@{}}
    \hline \hline
    & & & &  \\ [-2.8ex]  
    Cluster & {\it HST} & PI & Filter & Total Exp. & Data  \\
            & Program   &    &        & Time (sec) & Ref. \\ [0.3ex] \hline
    & & & &  \\ [-2.8ex]  
    NGC\,1651 & 12257 & L. Girardi    & F475W & \llap{1}440 & 1 \\
    NGC\,1850 & 11925 & S. Deustua    & F467M &  800 & 2 \\
              & 14174 & P. Goudfrooij & F814W &  790 & 2 \\
    NGC\,2173 & 12257 & L. Girardi    & F475W & \llap{1}520 & 1 \\
              &       &               & F814W & \llap{1}980 & 1 \\
    NGC\,2213 & 12257 & L. Girardi    & F475W & \llap{1}440 & 1 \\
              &       &               & F814W & \llap{1}520 & 1 \\
    NGC\,2249 & 12908 & P. Goudfrooij & F438W & \llap{1}650 & 3, 4 \\
              &       &               & F814W &  910 & 3, 4 \\ [0.3ex]  \hline
  \multicolumn{6}{@{}p{8cm}}{Data References: (1) \cite{Goudfrooij2014}, (2)
    \cite{Correnti17a}, (3) \cite{Correnti14}, (4)
    \cite{Goudfrooij17} }
  \end{tabular}
  \label{tab:HSTobs}
\end{table}

The results of the isochrone fitting are listed in Table~\ref{All_CMD} and depicted in the left-hand and middle panels of Figure~\ref{f:CMD_analysis}.
In general, we find that the quality of the fits of the MIST and Padova isochrones to cluster CMDs are very similar to each other for clusters younger than about 1 Gyr, when the RGB hasn't yet fully developed (see results for NGC 1850 and NGC 2249). For the remaining clusters, which have ages in the range 1.5\,--\,2 Gyr, we find that the relative positions of the MSTO and red clump in the CMD are better fit by the Padova isochrones than by the MIST ones. This seems to be mainly due to their different prescriptions for the treatment of convective overshoot (see \citealt{Marigo08} versus \citealt{Choi16}).

The population parameters derived from the MIST and Padova models are compared in  Figure~\ref{compare_CMD}. The derived ages are in good agreement between the two isochrone models, however the results of metallicity and reddening are not.

Finally, we also perform a 2-SSP fit to the distribution of stars across the MSTO of each cluster using Monte Carlo simulations, adopting the methodology of \citet[][see their Section 6.1]{Goudfrooij11a}. This was done using the Padova isochrones, since they provide a better fit to the MSTO morphology than the MIST models. These 2-SSP fits to the CMDs were done to enable a comparison to the 2-SSP fits to the integrated-light spectra of the eMSTO clusters (see Section~\ref{s:2-SSP fits} below).  The results of the 2-SSP fits are listed in Table~\ref{2SSP_CMD} and depicted in the right-hand panels of Figure~\ref{f:CMD_analysis}.

 \begin{table*}
\caption{Systematic CMD results for 1-SSP fits}
\label{All_CMD}
\begin{tabular}{@{}lllllllll@{}}
\hline \hline
Cluster & Age$^{1}$ & d(Age) & [Z/H] & d([Z/H]) & $E(B\!-\!V)$ & d$(E(B\!-\!V))$ & $m\!-\!M$ & d$(m\!-\!M)$ \\
\hline
MIST\\
NGC\,1651 & 9.32 & 0.03 & $-$0.50 & 0.05 & 0.10 &  0.01 & 18.37 & 0.03 \\
NGC\,1850 & 7.95 & 0.05 & $-$0.20 & 0.05 & 0.11 &  0.01 & 18.50 & 0.03 \\
NGC\,2173 & 9.27 & 0.03 & $-$0.50 & 0.05 & 0.09 &  0.01 & 18.37 & 0.03 \\
NGC\,2213 & 9.25 & 0.03 & $-$0.50 & 0.05 & 0.08 &  0.01 & 18.28 & 0.03 \\
NGC\,2249 & 9.00 & 0.03 & $-$0.30 & 0.05 & 0.01 &  0.01 & 18.23 & 0.03 \\
\hline
Padova\\
NGC\,1651 & 9.30 & 0.02 & $-$0.40 & 0.05 & 0.06 &  0.01 & 18.41 & 0.03 \\
NGC\,1850 & 7.95 & 0.03 & $-$0.32 & 0.05 & 0.11 &  0.01 & 18.50 & 0.03 \\
NGC\,2173 & 9.19 & 0.02 & $-$0.40 & 0.05 & 0.09 &  0.01 & 18.37 & 0.03 \\
NGC\,2213 & 9.23 & 0.02 & $-$0.40 & 0.05 & 0.05 &  0.01 & 18.36 & 0.03 \\
NGC\,2249 & 9.00 & 0.02 & $-$0.49 & 0.03 & 0.02 &  0.01 & 18.20 & 0.05 \\
%NGC\,1651 & 9.30 & 0.02 & 0.0080 & 0.001 & 0.19 &  0.02 & 18.41 & 0.03 \\
%NGC\,1850 & 7.95 & 0.03 & 0.0095 & 0.001 & 0.34 &  0.02 & 18.50 & 0.03 \\
%NGC\,2173 & 9.19 & 0.02 & 0.0080 & 0.001 & 0.28 &  0.02 & 18.37 & 0.03 \\
%NGC\,2213 & 9.23 & 0.02 & 0.0080 & 0.001 & 0.14 &  0.02 & 18.36 & 0.03 \\
%NGC\,2249 & 9.00 & 0.02 & 0.0065 & 0.0005 & 0.07 &  0.02 & 18.20 & 0.05 \\
\hline
\multicolumn{4}{l}{$^1$ log (Age/yr).}\\
\end{tabular}
\end{table*}

\begin{table}
\caption{Results from CMDs: 2-SSP fits}
\label{2SSP_CMD}
\begin{tabular}{@{}llll@{}}
\hline \hline
Cluster & log\,(Age(SSP$_1$)) & log\,(Age(SSP$_2$)) & $f^{1}_{\it SSP1}$ \\
\hline
NGC\,1651 & 9.26 & 9.30 & 0.32 \\
NGC\,1850 & 7.90 & 8.10 & 0.41 \\
NGC\,2173 & 9.18 & 9.24 & 0.60 \\
NGC\,2213 & 9.20 & 9.26 & 0.32 \\
NGC\,2249 & 8.98 & 9.08 & 0.59 \\
\hline
\multicolumn{4}{@{}l}{$^1$ The fractional contribution of SSP$_{1}$ by mass.}
\end{tabular}
\end{table}

\section{Spectral Analysis: Full Fitting Method}
\label{s:spectral_analysis}

\subsection{Our Sample}

We obtained the integrated spectra of our sample in the optical range with the Goodman spectrograph \citep{Goodman2004} on the 4 m SOAR telescope in the long-slit low resolution mode (1.03 arcsec wide slit; 600 gpm VPH grating; R=1500; 3600<$\lambda$<6250 Å) in December 2011 by scanning the cluster with the slit starting on the southern edge, with the slit aligned east-west \citep{Asad13}.
The average S/N per \AA\ for our sample is as follows: 50 for NGC\,1651, 250 for NGC\,1850, 100 for NGC\,2173, 45 for NGC\,2213 and 40 for NGC\,2249. Data reduction was performed using the generic longslit/IFU data reduction package implemented in IDL and described in detail in \cite{Chilingarian18}.

\subsection{Method}
\label{Traditional}

We use the flexible stellar population synthesis (FSPS) code \citep{Conroy09, Conroy10} operated through the Python package python-FSPS \citep{python-fsps} to produce a model grid that varies in both age (6.8 $<$ log\,(age) $<$ 10.2, in steps of 0.1 dex) and metallicity ($-$1.0 $<$ [Z$/$H] $<$ 0.2, in steps of 0.2 dex), with \citet{Kroupa01} IMF, for both MIST \citep{Choi17} and Padova \citep{Marigo08} isochrones with MILES spectral library \citep{MILESI, Vazdekis10, Vazdekis16}, which has a spectral resolution of 2.5 \AA. The selected metallicity range was guided by the fact that our star clusters are in the LMC whose metallicity is $\sim -0.4$.

We use the full-spectrum fitting technique by applying the $\chi^2$ minimization equation:
\begin{equation}
\chi^{2} = \sum_{\lambda=\lambda_{3700	A }}^{\lambda_{5000	A}} \frac{[(OF)_{\lambda} - (MF)_{\lambda}]^{2}}{(OF)_{\lambda_{4020.4	A}}}.
\label{Eq_1SSP}
\end{equation}
where OF is the observed flux and MF is the model flux, to obtain the best match between SSP models and the observed integrated spectra of clusters using the updated version of Analyzer of integrated Spectra for Age Determination (ASAD) \citep{ Asad13, Asad14} for the wavelength range $3700-5000$\,\AA\ to maximize sensitivity to age variations following the results of \citetalias{Asad20} (the subsequent findings of \cite{Goncalves20} were consistent with this). 

We convoluted the model spectra with a Gaussian to match the spectral resolution of the observations.
In equation \ref{Eq_1SSP}, we apply the \cite{Cardelli89} extinction law in the optical range (using $R_V$ = 3.1), using a range of $E(B-V)$ from 0.0 to 0.5, in steps of 0.01. While the age and reddening of a cluster are not physically correlated, they both affect the continuum shape of the spectrum in a similar way.

\subsection{Results}
\label{SSP1_results}

\begin{figure*}
\resizebox{150mm}{!}{\includegraphics[angle=0]{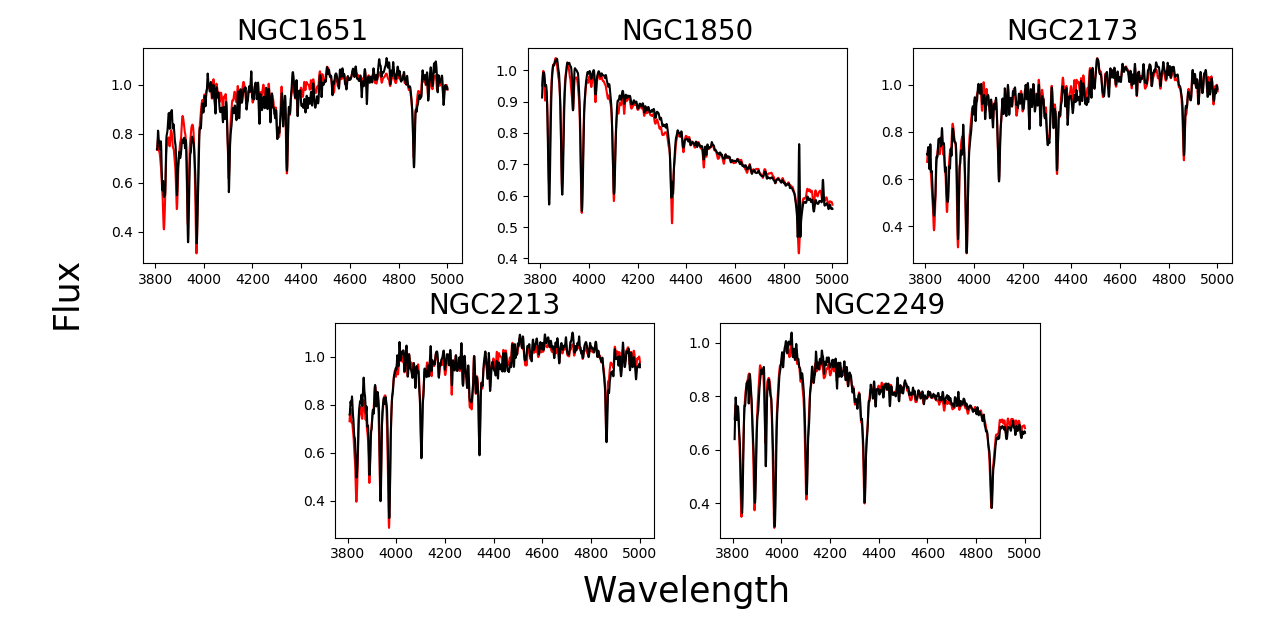}} \\
\caption{Best $\chi^2$ fits obtained with MIST isochrone models using the full-spectrum fitting technique. The observed spectrum is shown in black and the best matching model spectrum is shown in red.}
\label{MIST}
\end{figure*}

\begin{figure*}
\resizebox{150mm}{!}{\includegraphics[angle=0]{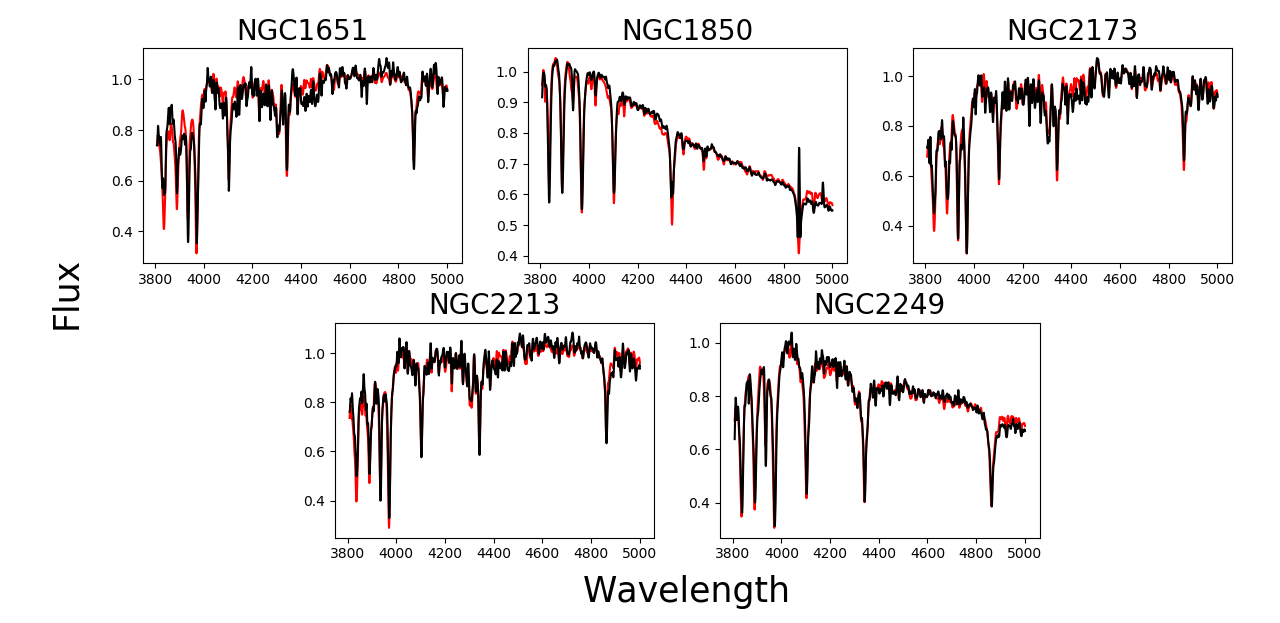}} \\
\caption{Best $\chi^2$ fits obtained with Padova isochrone models using the full-spectrum fitting technique. The observed spectrum is shown in black and the best matching model spectrum is shown in red.}
\label{Padova}
\end{figure*}

Figures \ref{MIST} and \ref{Padova} show the best $\chi^2$ fits obtained using the full-spectrum fitting technique with MIST and Padova isochrones respectively. There is no significant difference between the results obtained with MIST isochrones  versus those with Padova isochrones.

 \begin{table}
\caption{Results from integrated spectra, 1-SSP solution}
\label{Results_IntegratedSpectra}
\begin{tabular}{@{}llllll@{}}
\hline \hline
Cluster & Age$^{1}$ & d(Age) & [Z/H] & d([Z/H])& $E(B\!-\!V)$ \\
\hline
Padova\\
NGC\,1651 & 9.4 & 0.16 & $-$1.0 & 0.38 & 0.05 \\
NGC\,1850 & 7.5 & 0.03 & 0.0 & 0.30 & 0.09 \\
NGC\,2173 & 9.0 & 0.03 & 0.0 & 0.21 & 0.14 \\
NGC\,2213 & 9.4 & 0.16 & $-$1.0 & 0.38 & 0.07 \\
NGC\,2249 & 8.6 & 0.08 & 0.0 & 0.23 & 0.08 \\
\hline
MIST\\
NGC\,1651 & 9.5 & 0.17 & $-$1.0 & 0.38 & 0.04 \\
NGC\,1850 & 7.4 & 0.0 & 0.0 & 0.30 & 0.08 \\
NGC\,2173 & 9.1 & 0.17 & 0.0 & 0.21 & 0.11 \\
NGC\,2213 & 9.5 & 0.03 & $-$1.0 & 0.38 & 0.06 \\
NGC\,2249 & 8.6 & 0.08 & 0.0 & 0.23 & 0.08 \\
\hline
\multicolumn{4}{l}{$^1$ log (Age/yr).}\\
\end{tabular}
\end{table}

\begin{figure*}
\begin{tabular}{ccc}
\resizebox{90mm}{!}{\includegraphics[angle=0]{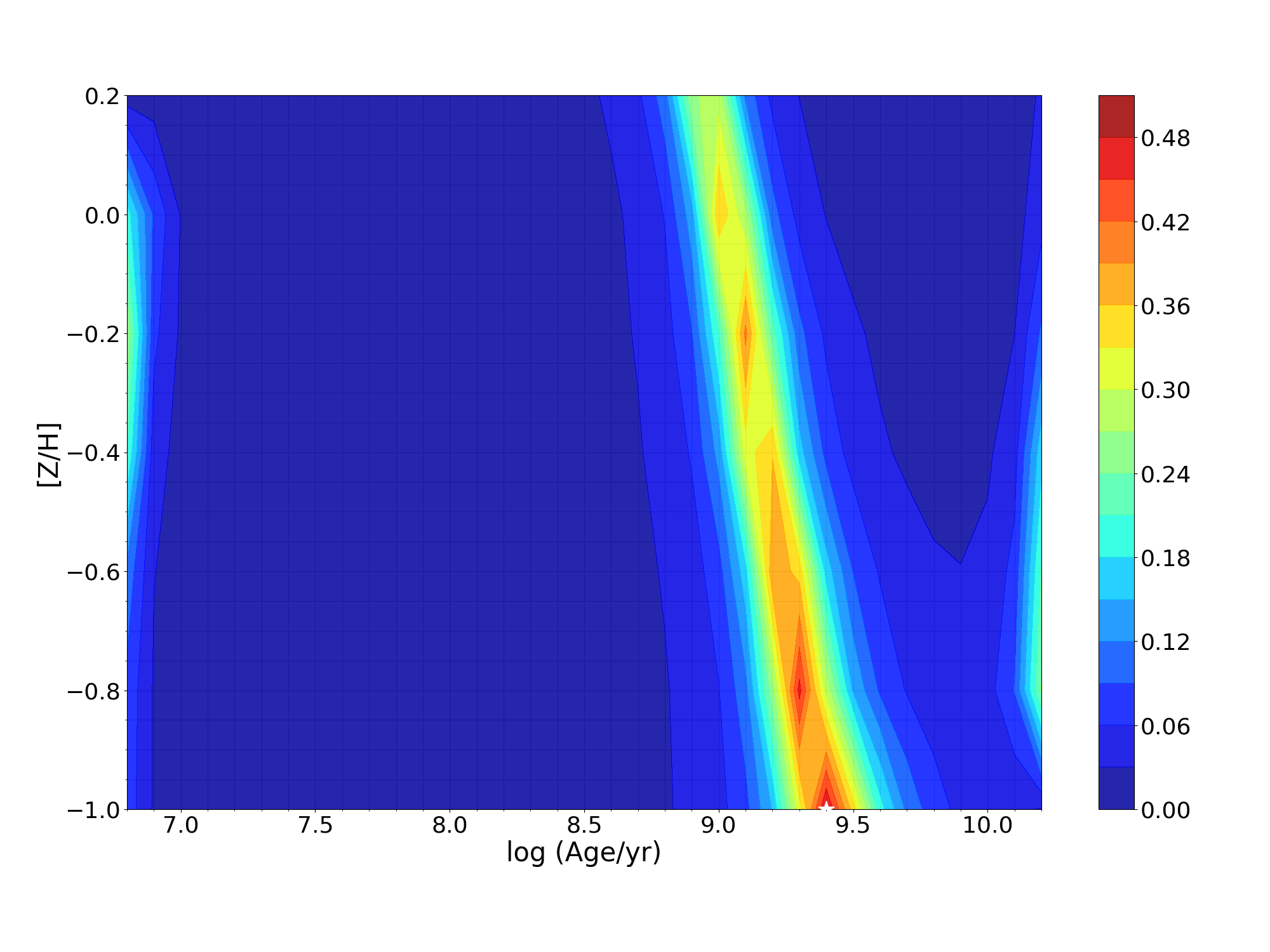}}
\resizebox{90mm}{!}{\includegraphics[angle=0]{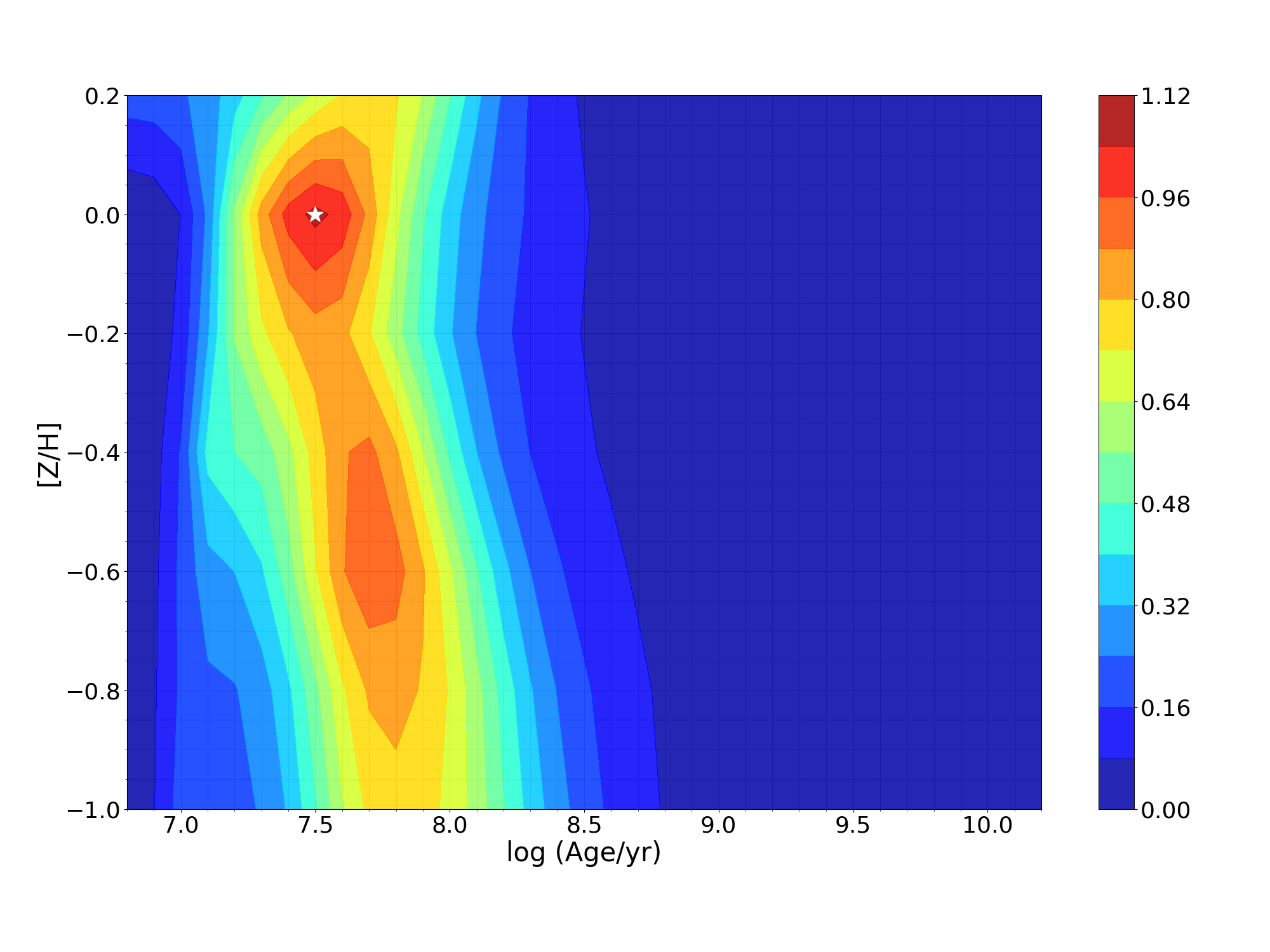}} \\
\resizebox{90mm}{!}{\includegraphics[angle=0]{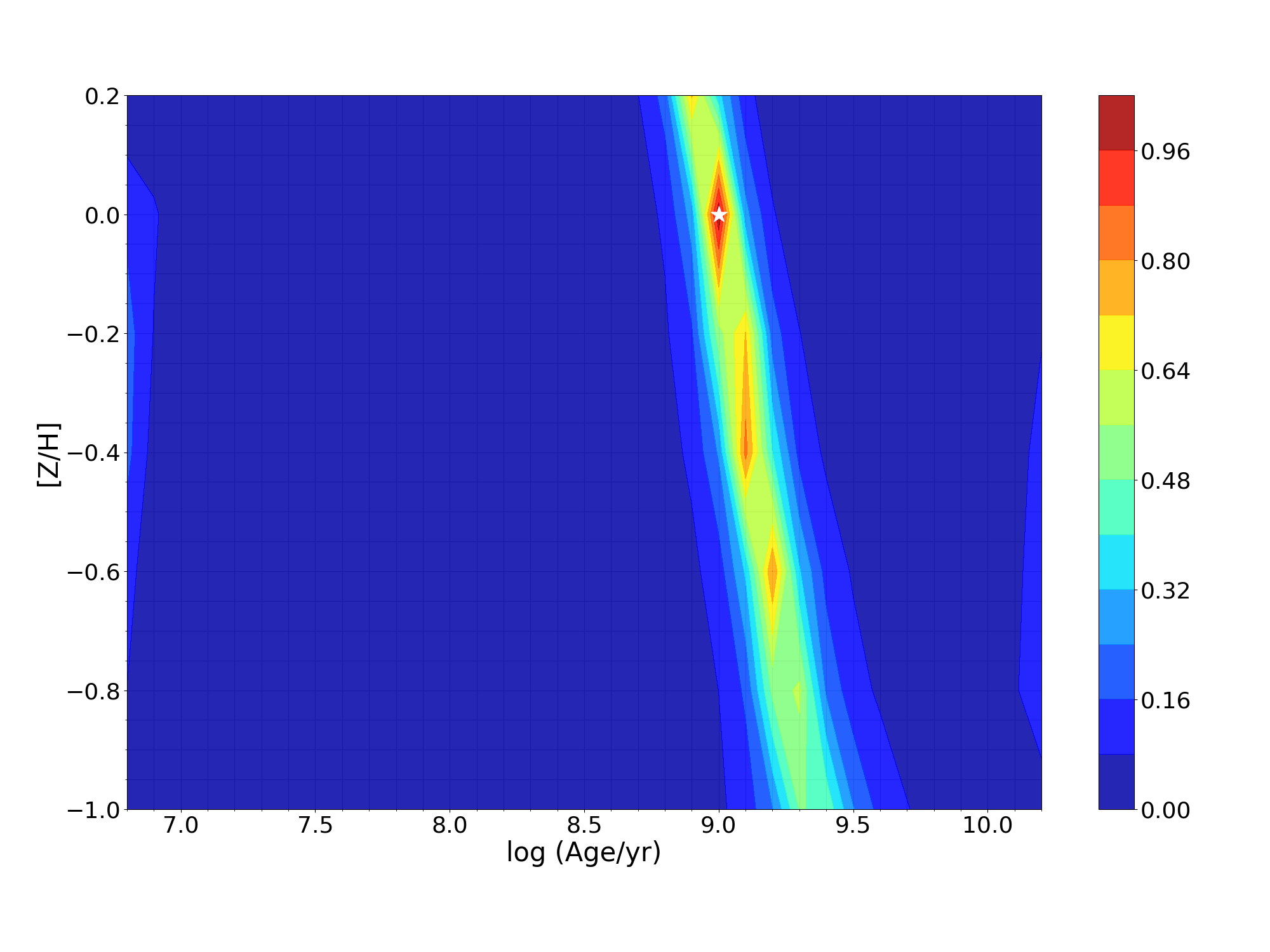}}
\resizebox{90mm}{!}{\includegraphics[angle=0]{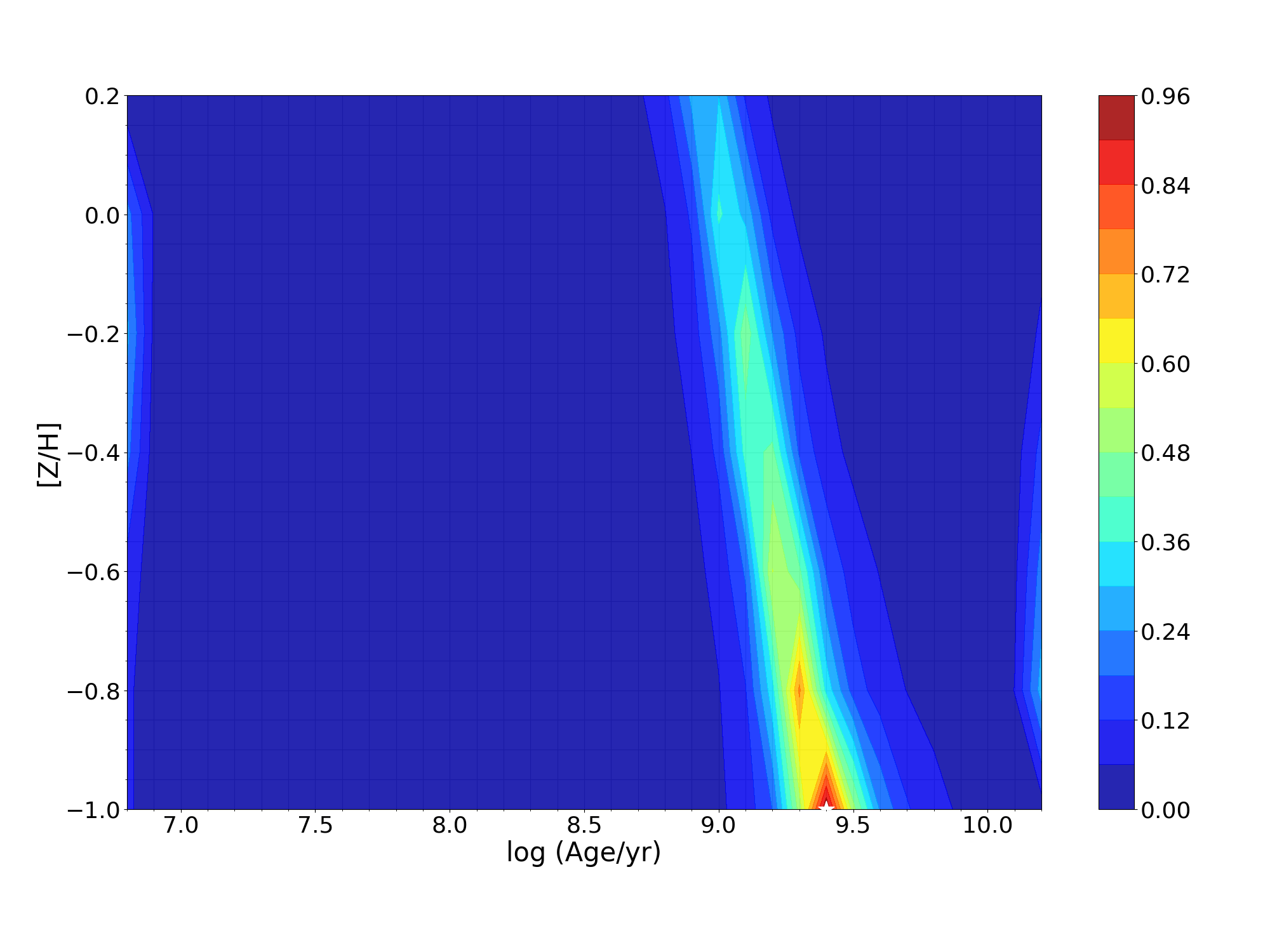}}\\
\resizebox{90mm}{!}{\includegraphics[angle=0]{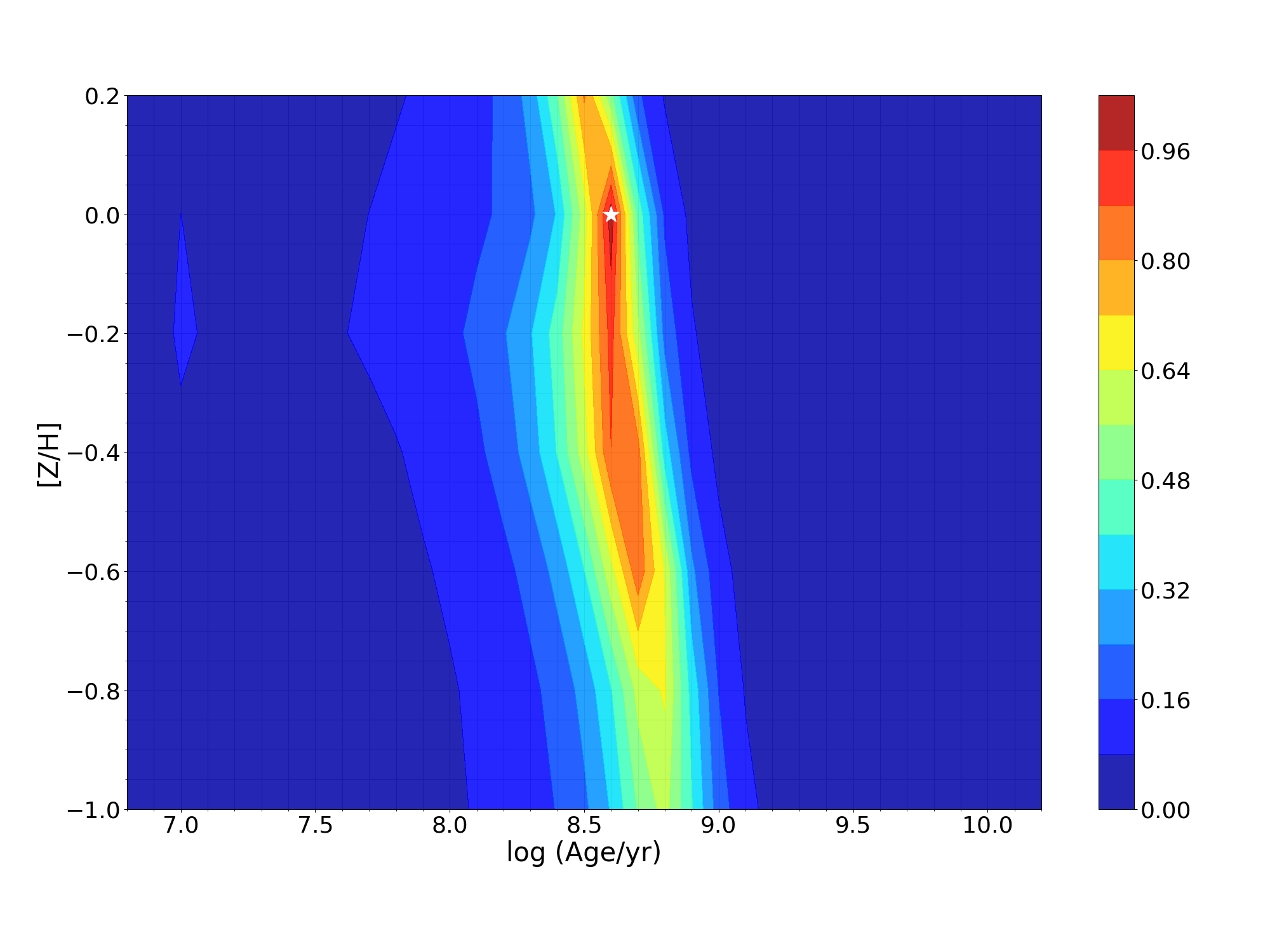}}\\
\end{tabular}
\caption{1/$\chi^2$ maps when using Padova isochrones.}
\label{chi2_inverse_Padova}
\end{figure*}

\begin{figure*}
\begin{tabular}{ccc}
\resizebox{90mm}{!}{\includegraphics[angle=0]{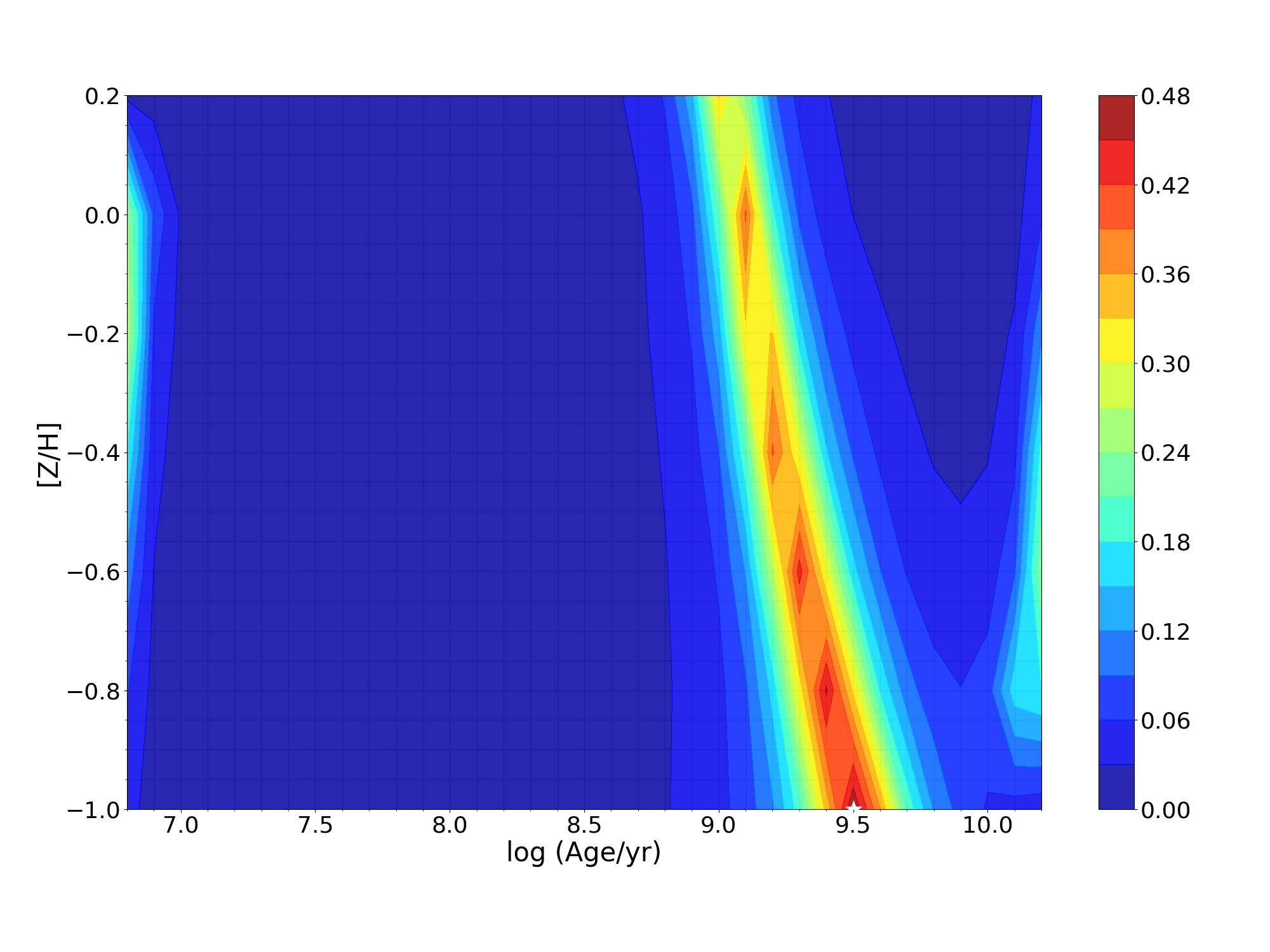}}
\resizebox{90mm}{!}{\includegraphics[angle=0]{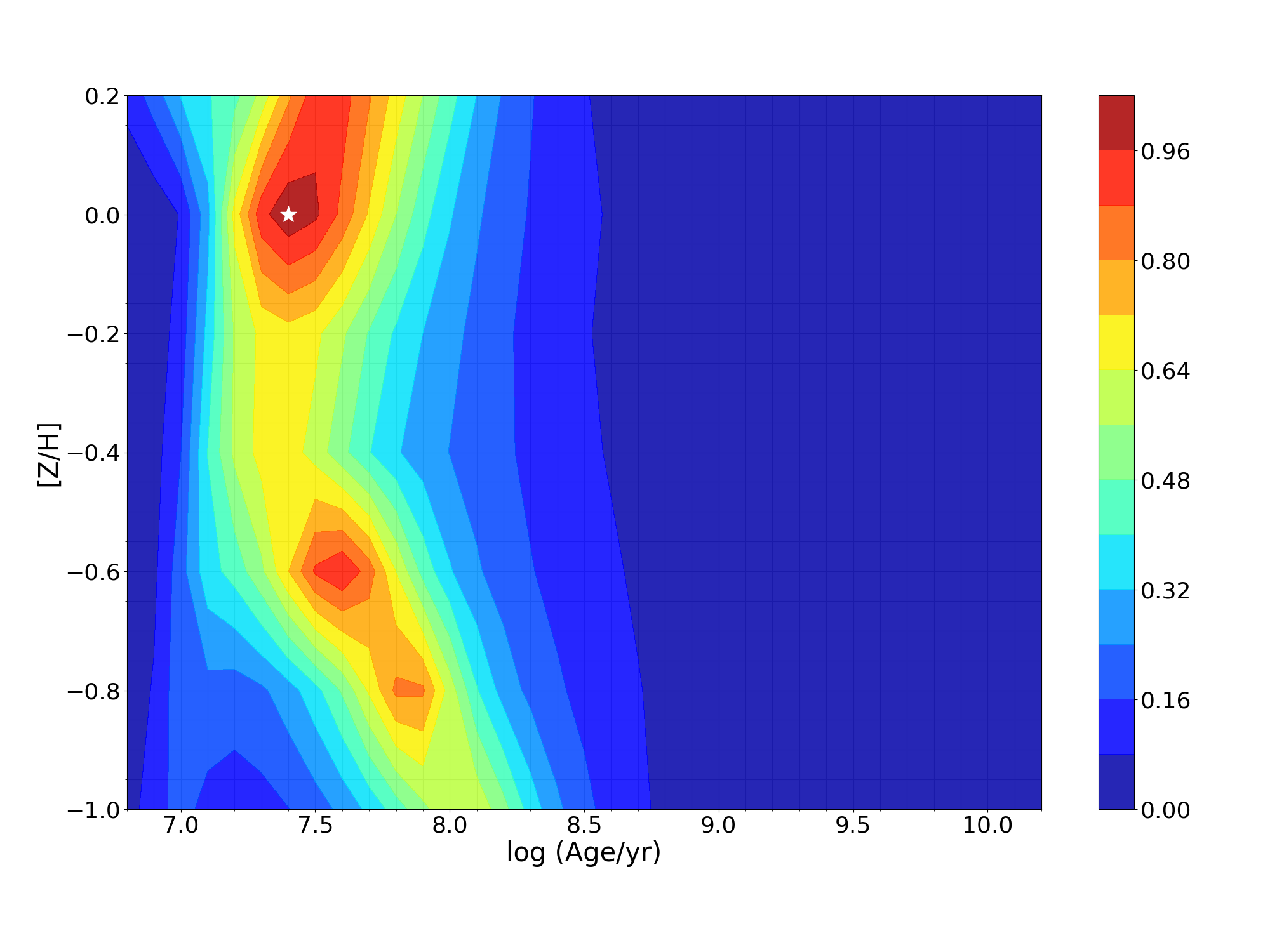}} \\
\resizebox{90mm}{!}{\includegraphics[angle=0]{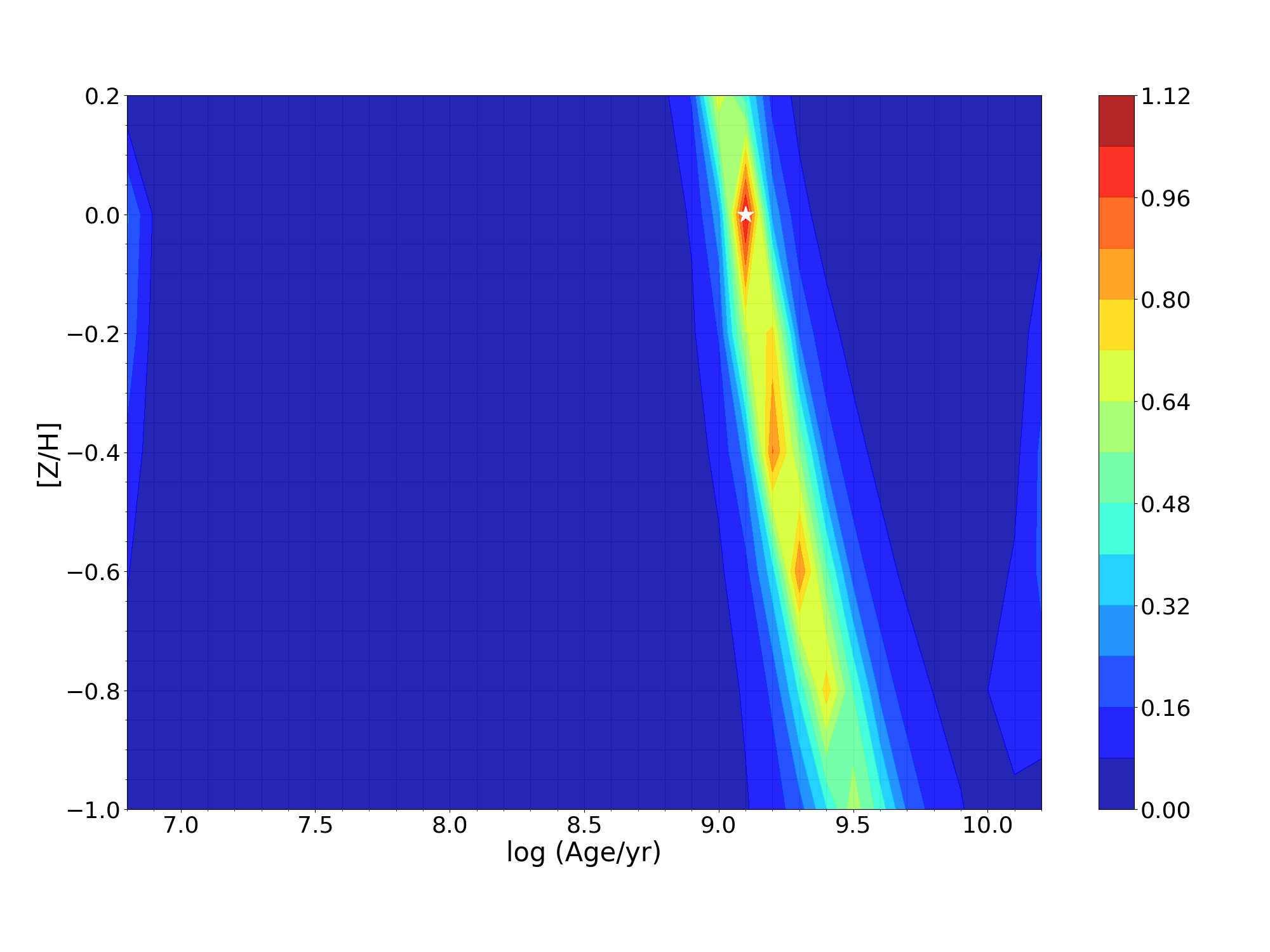}}
\resizebox{90mm}{!}{\includegraphics[angle=0]{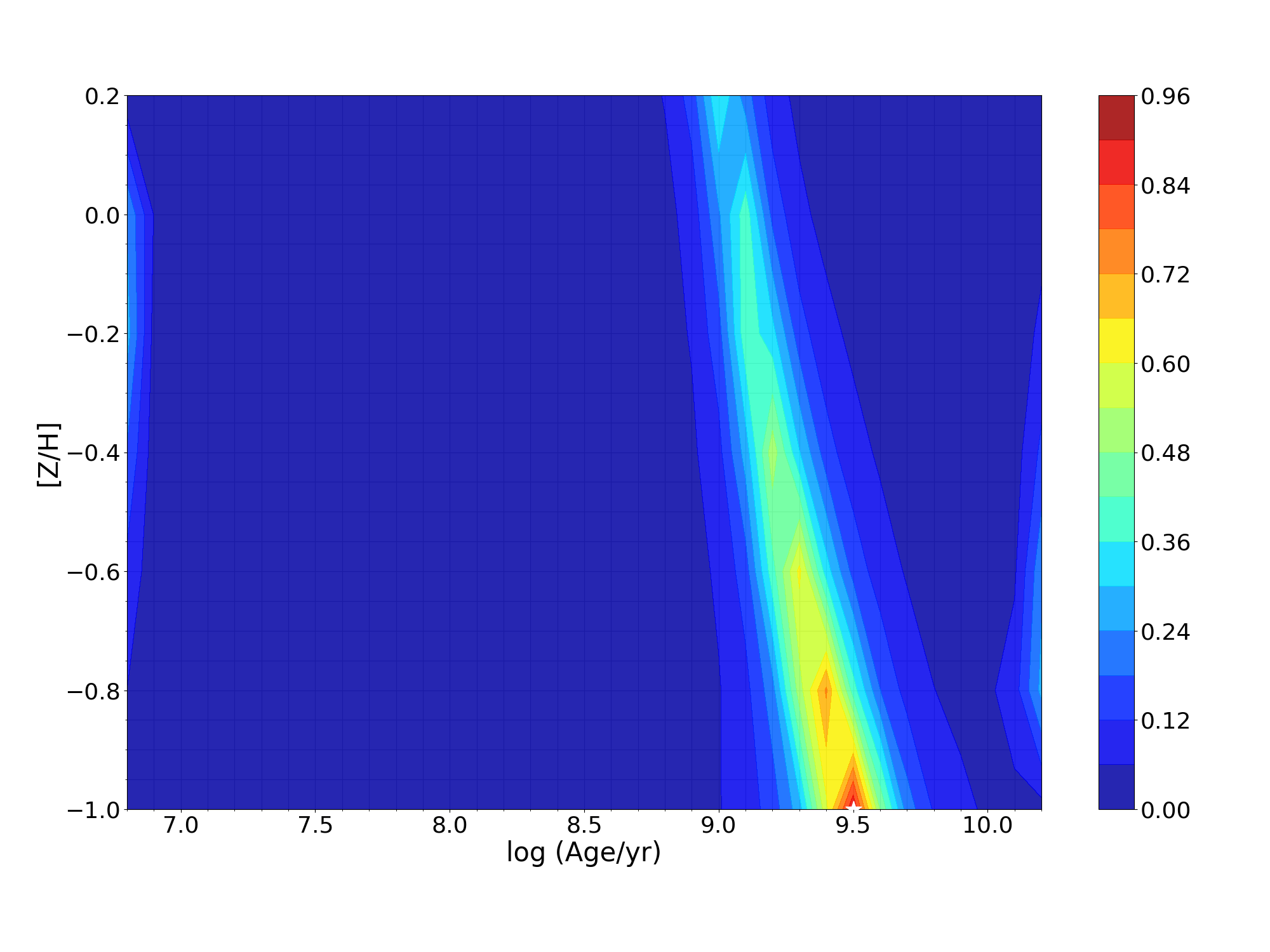}} \\
\resizebox{90mm}{!}{\includegraphics[angle=0]{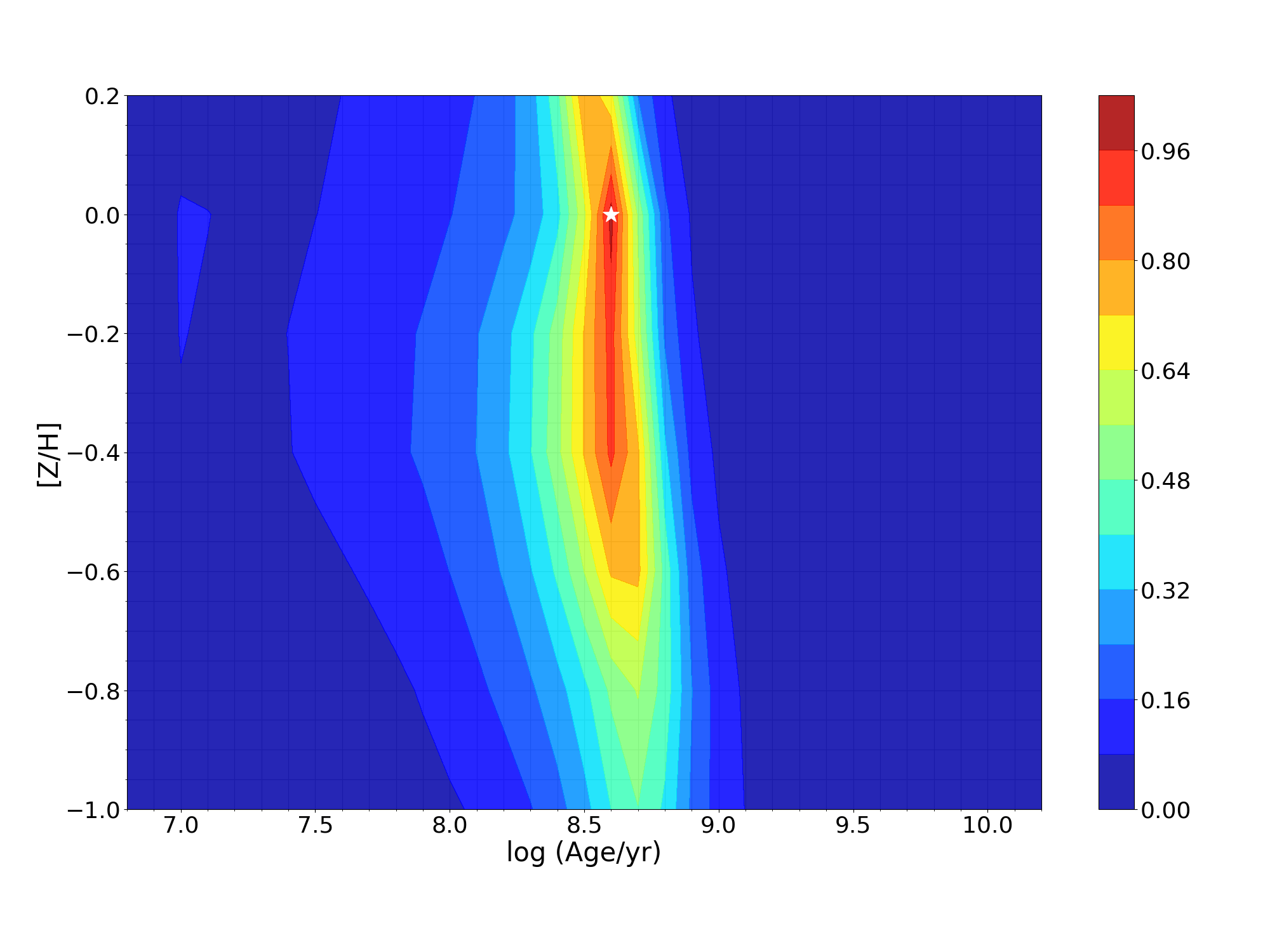}}\\
\end{tabular}
\caption{1/$\chi^2$ maps when using MIST isochrones.}
\label{chi2_inverse_MIST}
\end{figure*}

Figures \ref{chi2_inverse_Padova} and \ref{chi2_inverse_MIST} show the 2-D maps of $1/\chi^2$ when using the dereddened observed spectra. It is worth noting that in our analysis, reddening, age and metallicity were computed simultaneously in the same fit, however to make the 2-D plots in Figures \ref{chi2_inverse_Padova} and \ref{chi2_inverse_MIST}, we first corrected the observations for the reddening 
according to the value returned by the simultaneous fit and then fit again fixing the reddening.
The dark red region is the region with the lowest $\chi^2$ values, while the dark blue regions have the highest $\chi^2$. The white star indicates the location of the minimum $\chi^2$ value obtained by the fitting.  These maps provide information about the uniqueness of the solution (i.e., the age/metallicity values assigned for the cluster) as well as the precision of the result (a widely spread dark red region indicates a less precise result).
Overall, the map of each cluster does not change significantly when changing between MIST and Padova isochrone models. For almost all clusters, the maps show that the fitting result is more narrowly constrained for age than for  metallicity. In fact, the full-spectrum fitting procedure assigned the minimum [Z/H] = $-$1.0 for some clusters, along with large uncertainties. This is consistent with the simulations of \citetalias{Asad20} and \citetalias{Goudfrooij21} which showed that the full-spectrum fitting method can be used to obtain ages of SSPs to a precision of 0.1 dex in this wavelength range, while metallicity determinations are generally not as precise.

Figures \ref{chi2_inverse_Padova} and \ref{chi2_inverse_MIST} also illustrate the well-known age-metallicity degeneracy as a tilted band of low $\chi^2$ values.

\begin{figure*}
\resizebox{150mm}{!}{\includegraphics[angle=0]{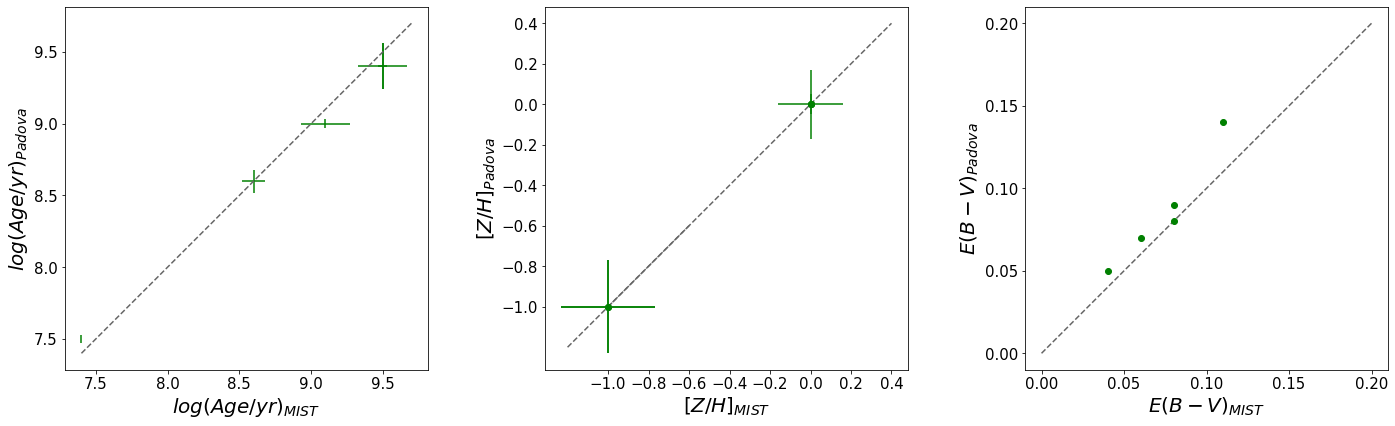}} \\
\caption{The correlations between the results derived using the full-spectrum fitting technique method for the Padova and MIST models respectively.}
\label{Compare_Spectra}
\end{figure*}

\begin{figure*}
\resizebox{150mm}{!}{\includegraphics[angle=0]{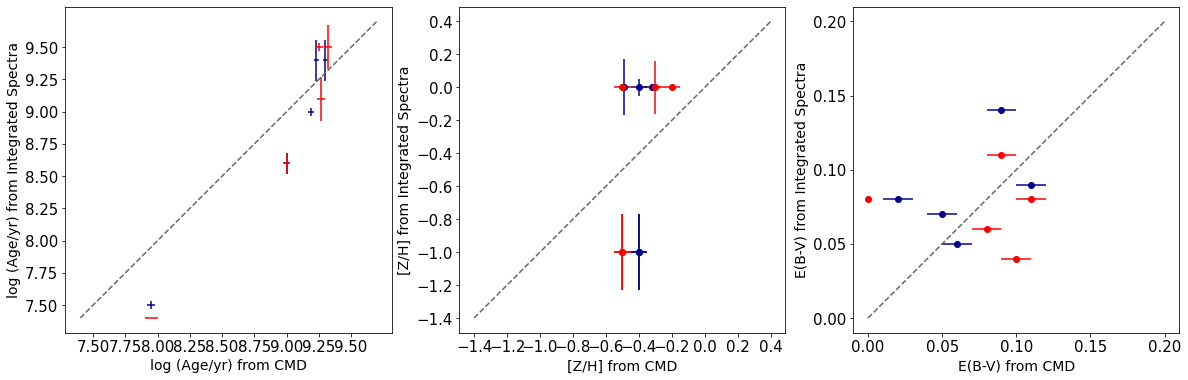}} \\
\caption{The correlations between the results derived using the full-spectrum fitting technique and those using the CMD method. Results obtained using MIST isochrones are shown in red and those obtained using Padova isochrones are shown in blue. }
\label{Final_Results}
\end{figure*}

The ages, metallicities and reddening values derived for our sample are listed in table \ref{Results_IntegratedSpectra} and illustrated in Figure \ref{Compare_Spectra}. The uncertainties listed in Table \ref{Results_IntegratedSpectra} are derived from the results of \citetalias{Asad20} and \citetalias{Goudfrooij21} who investigated the precision of the ages and metallicities of 21,000 mock simple stellar populations (SSPs) as derived using the full-spectrum fitting method. These uncertainties reflect the standard deviation of the difference between the derived and the true parameter value for each input age, S/N, and cluster mass in this wavelength region. The derived results for age and metallicity are in good agreement between the two isochrone families, while the reddening results agree to within $\pm$0.01 mag in $E(B\!-\!V)$ for all clusters except NGC\,2173.
Figure \ref{Final_Results} shows the correlation between the parameters obtained from full-spectrum fitting of integrated spectra versus the ones obtained from CMDs. Results obtained using MIST and Padova isochrones are shown in red and blue, respectively. The derived ages using the two methods are generally in good agreement, except for the case of the young cluster NGC\,1850.
It is worth mentioning that for NGC\,1850 the main cluster is NGC\,1850a, which has an age of $\sim90$~Myr (e.g., \citep{Bastian16}).  However, there is a younger cluster ($\sim$\,10 Myr) named NGC 1850b enclosed within NGC\,1850. NGC\,1850b is much less massive, but contains a few bright O stars.  The two clusters do not appear to be related, with NGC\,1850b being located behind NGC\,1850a. NGC\,1850b is included in our slit \citep{Asad13}, so this is a case where we might expect to find a young population that makes up $\sim$\,5-10\% of the mass.
The [Z/H] and reddening values derived from the two methods are found to generally yield different results, without a clear hint of a trend with age, metallicity or reddening. A larger sample is needed in order to detect and/or analyze any such trends.

\subsection{Fitting 2-SSP to Each Cluster}
\label{s:2-SSP fits}

In this Section, we probe the ability of the full-spectrum fitting method to detect age spreads in eMSTO clusters such as those in our sample. As such, we combine two SSPs to obtain the best fit between the models and our clusters following the method introduced in \citetalias{Asad21}. From the model described in Section \ref{Traditional}, we select the grid for the age range 7.0 $<$ log age $<$ 10.2 in steps of 0.1 dex. We use a fixed metallicity [Z$/$H] = $-$0.4 in this context, since the CMD studies found no evidence for spreads in metallicity in eMSTO clusters (see Section~\ref{s:CMDanalysis} and references therein). We use Padova isochrones in order to compare the results with those obtained from CMDs in section \ref{s:CMDanalysis}.

We then combine two SSPs according to the equation:
\begin{equation}
f\,{\rm SSP}_{1}+(1-f)\,{\rm SSP}_{2}
\label{Eq_2SSPs}
\end{equation}
\noindent where $f$ is the fractional contribution of SSP$_{1}$ by mass, running from 0 to 1 in steps of 0.1, SSP$_{1}$ is the SSP of an arbitrary age, and SSP$_{2}$ is the additional SSP.
Results are listed in Table \ref{2SSP_IntegratedSpectra}.

 \begin{table}
\caption{Results from integrated spectra, 2-SSP solution}
\label{2SSP_IntegratedSpectra}
\begin{tabular}{lllllllll}
\hline \hline
Cluster & log\,(Age(SSP$_1$)) & log\,(Age(SSP$_2$)) & $f^{1}$ \\
\hline
NGC\,1651 & 9.5 & - & 1.0 \\
NGC\,1850 & 7.0 & 8.0 & 0.1 \\
NGC\,2173 & 9.0 & 9.2 & 0.2 \\
NGC\,2213 & 9.5 & - & 1.0 \\
NGC\,2249 & 8.5 & - & 1.0 \\
\hline
\multicolumn{4}{l}{$^1$ The fractional contribution of SSP$_{1}$ by mass.}
\end{tabular}
\end{table}

As discussed in \citetalias{Asad20}, the success rate (defined as the number of times the derived parameter value matches the input value $\pm$ 0.1 dex) peaks around log (age/yr) = 7.2 and 9.0 and drops around 8.6 and $\ga$ 9.5 because SSP SEDs are very similar around those ages, which renders it difficult to identify their correct age. This explains at least in part why our fitting algorithm returns a 1-SSP solution for three eMSTO clusters in our sample (NGC\,1651, NGC\,2213 and NGC\,2249) with log (age/yr) = 9.5, 9.5, and 8.5, respectively. This result calls into question the claim by \citet{Cabrera-Ziri16} who found no evidence for an age spread in an integrated spectrum of the massive cluster W3 in the merger remnant galaxy NGC\,7252 with log (age/yr) $\sim$ 8.6, and then argued that this rules out an extended star formation history like those inferred for eMSTO clusters.

\begin{figure*}
\begin{tabular}{ccc}
\resizebox{75mm}{!}{\includegraphics[angle=0]{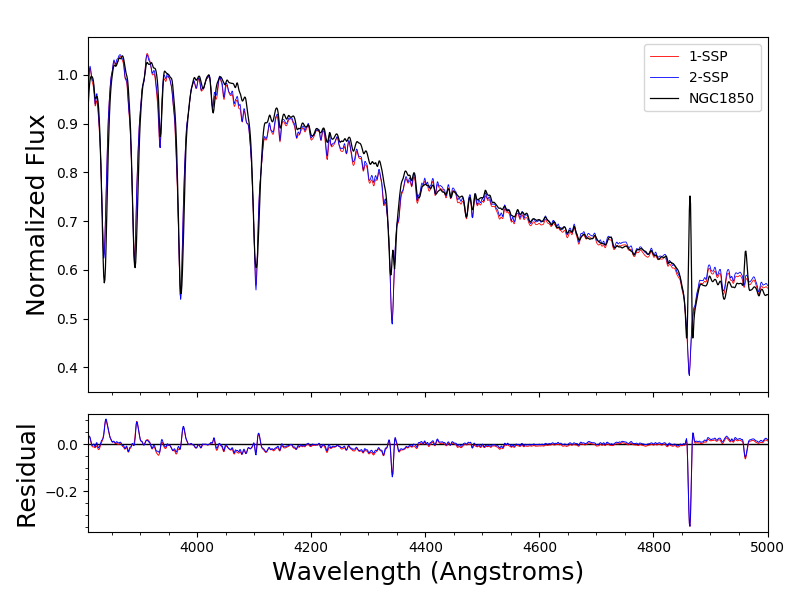}}
\resizebox{75mm}{!}{\includegraphics[angle=0]{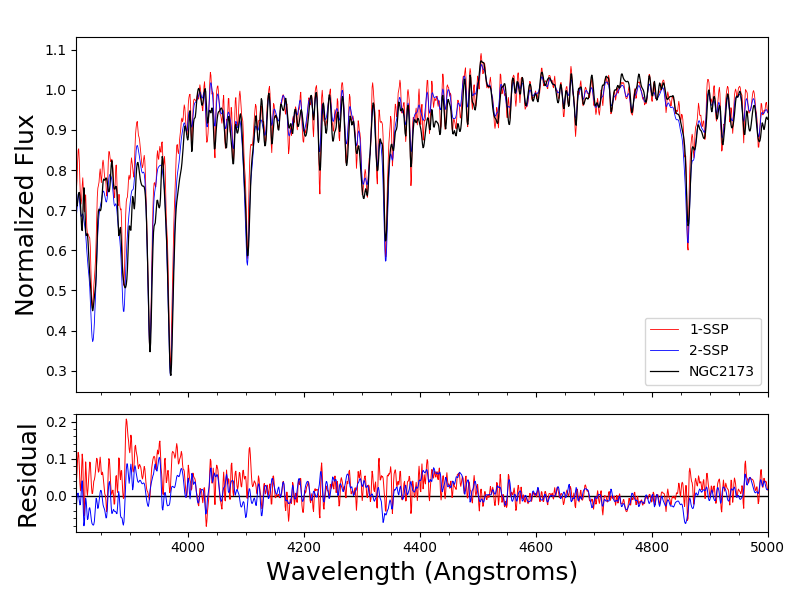}} \\
\end{tabular}
\caption{Best $\chi^2$ fits with SSP models compared to the best fits obtained with 2-SSP combinations.}
\label{2SSP_plots}
\end{figure*}

Figure \ref{2SSP_plots} shows the best $\chi^2$ fits with SSP models compared to the best fits obtained with 2-SSP combinations for the two clusters NGC\,1850 and NGC\,2173 for which a 2-SSP solution is predicted in Table \ref{2SSP_IntegratedSpectra}. As discussed in Section \ref{SSP1_results}, NGC\,1850b is included in our slit for the NGC\,1850 integrated spectrum \citep{Asad13}, which is likely why the 2-SSP solution is identifying two populations, one of log (age/yr) = 7.0 and the other with log (age/yr) = 8.0.The fact that the full-spectrum fitting technique
is able to identify these two populations confirms the good performance of the method.
We notice an H$\beta$ emission line in the NGC\,1850 spectrum. It is worth mentioning that no emission lines or nebular continuum were included when we generated the grids of FSPS models; hence, the inference of a young subpopulation by our 2-SSP fit for NGC\,1850 must be mainly due to the shape of the stellar continuum and absorption lines. We will investigate the effect of emission lines and nebular continuum more closely in future work on younger star clusters and complex stellar populations in galaxies.

\section{Summary and Concluding Remarks}
\label{Summary}

We provide the first systematic comparison between ages obtained from CMDs and integrated spectra for five star clusters in the LMC. Our results are summarized in the following points:\\
A) The ages derived using Padova isochrones match those derived using MIST isochrones for both resolved and resolved data. This is consistent with \citetalias{Asad20} who showed that significant differences between the two isochrone families only occur in the age range $8.0 \la \mbox{log\,(age/yr)} \la 8.5$ for the wavelength range considered here; however, our sample does not have clusters in this age range. \\
B) The metallicity derived using Padova isochrones is often significantly different from the metallicity derived using MIST isochrones when using the CMD method. However, they match perfectly when using the full-spectrum fitting method.\\
C) The reddening derived using Padova isochrones match the reddening derived using MIST isochrones using the CMD method for two of of five clusters. The other three values are underestimated by Padova isochrones compared to MIST isochrones. However, when using the full-spectrum fitting method, the results obtained by Padova isochrones tend to be slightly \emph{over}estimated compared to those obtained using MIST isochrones.\\
D) In terms of age determination, there is a good agreement between the ages derived from CMDs and integrated spectra. \\
E) The metallicity derived from full-spectrum fitting generally does not match that derived from CMDs. This is likely due to the relative lack of precision of metallicity determinations using the full-spectrum fitting technique for the masses and ages of these clusters. Combining the results of \citetalias{Asad20} and \citetalias{Goudfrooij21}, we find that for a typical cluster mass of $3\times10^4\;M_{\odot}$ and a spectrum with S/N $\geq$ 50/\AA, typical uncertainties for [Z/H] range between $\sim$\,0.20 and 0.45 dex, depending on the age.
\\
F) Using 2-SSP fits to the integrated spectra, we obtain single-age results for three eMSTO clusters with log (age/yr) = 8.5 and 9.5. This is consistent with the finding of \citetalias{Asad20} that the success rate of full-spectrum fitting is low around 8.6 and $>$ 9.5 because SSP SEDs are very similar around those ages. This insensitivity to age spreads should be taken into account in spectroscopic studies of age spreads in intermediate-age clusters (which typically feature eMSTOs in CMDs). \\
G) The $\sim$\,90 Myr old cluster NGC\,1850, which has a young, low-mass cluster (age $< 10$ Myr) superimposed on the sky, is an interesting example showing how well integrated-light spectroscopy can identify populations of largely different ages.  The $\chi^2$ distribution of our fits show a clear bimodality, reflecting the existence of two populations with two different ages and metallicities.

\section*{Acknowledgements}

RA thanks the Space Telescope Science Institute for a sabbatical visitorship including travel and subsistence support as well as access to their science cluster computer facilities. \\
RA is grateful for Nate Bastian and Silvia Martocchia for the useful discussions.\\
This work is supported by the FRG grants P.I., R.\ Asa'd from American University of Sharjah.

\section*{Data availability}

The spectra presented in this work will be made available. 

\bibliographystyle{mnras}
\bibliography{references}

\end{document}